\documentclass[titlepage,a4paper,11pt]{article}

\usepackage[dvips]{graphicx}

\usepackage[myheadings]{fullpage}
\usepackage{pmetrika}
\usepackage{fancyvrb}
\usepackage{verbatim}
\usepackage{pmbib}
\usepackage{pdflscape}
\usepackage{submit}
\usepackage{booktabs}
\usepackage{longtable}

\usepackage{rotating}

\usepackage{caption}
\usepackage{hyperref}
\usepackage{subfig}
\usepackage{amsmath}
\usepackage{amsfonts}
\usepackage{natbib}
\usepackage{multirow}
\usepackage{ragged2e}
\usepackage{tikz}
\usepackage{adjustbox}
\usetikzlibrary{arrows,shapes}
\usetikzlibrary{shapes,decorations,arrows,calc,arrows.meta,fit,positioning}
\tikzset{
	-Latex,auto,node distance =1 cm and 1 cm,semithick,
	intervention/.style ={rectangle, draw=black},
	random/.style ={circle, draw=black, fill=gray,
		fill opacity = 0.3,inner sep=0pt,
		text opacity=1, minimum size=1cm,},
	parameter/.style ={diamond, draw=black},
}

\makeatletter
\newcommand*\bigcdot{\mathpalette\bigcdot@{.75}}
\newcommand*\bigcdot@[2]{\mathbin{\vcenter{\hbox{\scalebox{#2}{$\m@th#1\bullet$}}}}}
\makeatother

\begin{document}

\begin{titlepage}

\title{Nodal heterogeneity can induce ghost triadic effects in relational event models}


\author{R{\=u}ta Juozaitien{\. e}*}
\affil{Vytautas Magnus university, Kaunas, Lithuania}

\author{Ernst C. Wit}
\affil{Università della Svizzera italiana, Lugano}

\vspace{\fill}\centerline{\today}\vspace{\fill}

\linespacing{1}
\contact{Correspondence should be sent to\\

\noindent E-Mail: ruta.juozaitiene@vdu.lt}

\end{titlepage}

\setcounter{page}{2}
\vspace*{2\baselineskip}

\RepeatTitle{Nodal heterogeneity can induce ghost triadic effects in relational event models}\vskip3pt

\linespacing{2}
\justifying
\abstracthead
\begin{abstract}
Temporal network data is often encoded as time-stamped interaction events between senders and receivers, such as co-authoring scientific articles or communication via email. A number of relational event frameworks have been proposed to address specific issues raised by complex temporal dependencies. These models attempt to quantify how individual behaviour, endogenous and exogenous factors, as well as interactions with other individuals modify the network dynamics over time. It is often of interest to determine whether changes in the network can be attributed to endogenous mechanisms reflecting natural relational tendencies, such as reciprocity or triadic effects. 

The propensity to form or receive ties can also, at least partially, be related to actor attributes. Nodal heterogeneity in the network is often modelled by including actor-specific or dyadic covariates. However, comprehensively capturing all personality traits is difficult in practice, if not impossible. A failure to account for heterogeneity may confound the substantive effect of key variables of interest. This work shows that failing to account for node level sender and receiver effects can induce ghost triadic effects. We propose a random-effect extension of the relational event model to deal with these problems. We show that it is often effective over more traditional approaches, such as in-degree and out-degree statistics. These results that the violation of the hierarchy principle due to insufficient information about nodal heterogeneity can be resolved by including random effects in the relational event model as a standard.  

\end{abstract}\vspace{\fill}\pagebreak

\section{Introduction}
\label{intro}

Social relationships are shaped by individuals' interpersonal actions, which play a crucial role in both forming and maintaining these connections \citep{borgatti2011network}. According to \cite{hinde}, relationships can be defined as series of interactions over time. Research on social interactions highlights that exchange participants influence each other's behavior \citep{raush1965interaction}. For instance, when we greet someone, we typically expect and receive a greeting in return. Similarly, in the context of email communication, when we send an inquiry or request, we anticipate and hope for a response, demonstrating the expectation of reciprocity in social exchanges. The formation of social ties is also strongly influenced by homophily, which refers to the tendency of individuals to prefer interacting with others of similar type \citep{mcpherson2001birds}.

Another commonly observed feature of social interactions is the tendency to form more complex closed structures. In its simplest manifestation these are triads. This mechanism assumes that new connections frequently emerge between people sharing common acquaintances. The results of several independent studies suggest that triadic closure can be identified as one of the fundamental dynamical principles in network formation and evolution \citep{Li,Klimek,Leskovec:2008}. Moreover, this tendency is widely supported on empirical grounds, since it can explain salient features of empirical social networks, including a strong community structure, fat-tailed degree distributions and high clustering coefficients \citep{foster,PhysRevE,PhysRevLett,model_added_nodes}. 

Researchers in both personality and social psychology acknowledge that also personality differences influence social relationships \citep{back2015opening,geukes2019explaining}. Individuals' unique personality traits and characteristics can shape how they interact with others. Due to the heterogeneity in their expansiveness, some actors tend to make many connections, while others prefer to stay on their own. \emph{Expansiveness} represents a person's degree of sociability and how much they enjoy being in crowds. Expansive people usually have a lower threshold for friendship, and as a consequence, they consider more people as their friends \citep{olk2010dynamics}. A closely related concept, representing the tendency to receive interactions, is \emph{popularity}. This feature reflects other people's attitudes towards a particular person. Popular individuals tend to be more often the receiver of relation events. Although both expansiveness and popularity might be functions of other underlying traits, such as genetics or status, respectively, in empirical studies those traits may not be recorded. Thus, social interactions are influenced by a complex interplay of individual characteristics, environmental context, and the history of past interactions. 

Relational event modeling \citep{butts2023relational} provides a flexible approach to studying the dynamic nature of social relationships. This framework attempts to quantify how individual behaviour, external factors and interaction with other individuals change the social network structure over time. It is often of interest to determine whether changes in the network can be attributed to endogenous mechanisms reflecting natural relational tendencies, such as reciprocity or triadic effects. Nodal heterogeneity in the network is often modelled by including actor-specific or dyadic covariates, such as age, gender, age difference, etc. However, capturing the full extend of all personality traits that encompass popularity or expansiveness is often difficult, if not impossible. 

This problem has also been encountered in other areas of network modeling. It has led to the development of random-effects models accounting for the latent and nodal heterogeneity \citep{thiemichen2016,BoxSteffensmeier2019,box2018modeling,kevork2021iterative}. In most cases, the individual levels of the random nodal effects are not of interest. However, accounting for additional heterogeneity is important to avoid bias in the estimation of other effects. Thus the inclusion of random effects is an elegant and straightforward way to handle the problem of an increasing number of parameters with an increasing number of actors. And, more importantly, this approach allows us to account for heterogeneity that may have significant implications for statistical network modelling and inference. Alternatively, various endogenous statistics such as nodal in-degree and out-degree can be used to reduce nodal heterogeneity. The interpretation of these effects, however, are rather different. Whereas random effects suggest the existence of unmeasured traits that are responsible for the network dynamics, nodal degree statistics effects imply the existence of emerging viral dynamics in the network.

Exponential random graph models (ERGMs) are a class of statistical models often used for modelling social networks. These models aim to identify features that explain the global structure of a network. A well-known issue in ERGMs is that a failure to account for heterogeneity may confound the substantive effect of key variables of interest. It has been shown \citep{thiemichen2016} that triadic closure estimates obtained using the model ignoring heterogeneity can vastly overstate the triadic effect present in the network. For example, if specific individuals are more outgoing than others, ERGMs not accounting for heterogeneity may confuse this feature with a network tendency towards a triadic closure \citep{BoxSteffensmeier2019}. These results suggest that the existence of heterogeneity may effect the conclusions drawn from ERGMs fitted on real world networks. 

The problem that omitting individual-level predictors of tie formation can bias parameters of endogenous network parameters, like reciprocity or transitivity, is well-known. Among network modellers, it is generally known as the ``hierarchy principle,'' that is, include underlying substructures in the model when modelling more complicated dynamics. This is especially true for ERGMs \citep{lusher2013exponential} and SAOMs \citep{snijders2017stochastic}, but also various applications of relational event models discuss the importance of a proper representation of degree dynamics in order to obtain credible parameter estimates of transitive closure \citep{Snijders2010, corbo2016new}. Usually, it is recommended to model, at least, in-degree and out-degree centralisation using in-stars and out-stars (or geometrically weighted versions of them) to represent degree dynamics that are not well captured by exogenous variables. The advantage of modelling nodal heterogeneity using endogenous star-parameters is that it allows to make conclusions about emergent degree dynamics: it predicts that the network dynamics alone is responsible for emerging patters. Exogenous variables, by their very definition, do not describe emergence, but relate the changes of the network dynamic to some external process. Nevertheless, sociological processes are well-known for their overdispersion, that is, presenting more individual level variability than is possible to model parsimoniously. 

In relational event modeling, various other approaches have been proposed to account for the node heterogeneity. \citep{butts} proposed including for each individual a fixed effect defined as a standard indicator function. The corresponding parameters then represent logged rate multipliers for all events having the corresponding individuals as senders or receivers. This can increase the number of parameters dramatically, and it does not distinguish between expansiveness or popularity effects. Other approaches include stochastic blockmodeling, which assumes latent groups of individuals having similar interaction tendencies \citep{dubois2013stochastic}, or dynamic latent space relational event modeling, which allows individuals' interactions to depend on dynamic locations in a latent space \citep{Igor2023}. 

Although it is standard to include individual level random effects in most sociological statistical models, this has only recently been introduced for relational event models \citep{uzaheta2023random}. As it is clear that the hierarchy principle is an important modelling concept, we propose adding individual node-level frailty terms as a general modelling strategy. Relational event models focus on behavioural interactions, which are defined as discrete events connecting a sender and a recipient at a specific point in time. In this manuscript we aim to show how node level popularity in terms of sender and receiver effects may mask ghost triadic effects. We propose a frailty model for reciprocal and triadic effects in relational event networks to disentangle them from node-specific effects such as popularity and expansiveness. 

\section{Relational event models with frailty}


The basic idea of relational event models involves modelling the evolution of social interactions as the outcome of a stochastic point process. A general framework capable of exploiting the information contained in sequences of relational events has been introduced in \cite{butts}. The model assumes time-stamped network data consisting of sequences $E = \{e_1, e_2,\ldots,e_T\}$ of relational events $e=(i,j,t)$, encoding the event time $t$, the event sender $i$ and the event receiver $j$. The events in this series are typically dependent on each other, as relational events often trigger others, such as, e.g., replying to a message or turn-taking in conversations. This interdependence is indeed one of the main interests of network analysis, since it can identify endogenous and exogenous drivers of how people interact \citep{Stadtfeld_2017}.

The relational event model assumes that every interaction process can be encoded by a multivariate counting measure. Following \citet{Wolfe}, a counting process for the directed edge between sender $i$ and receiver $j$ is defined as:
\begin{equation*}
	N_{ij}(t) = \# \{\text{relational events } i \rightarrow j \text{ up to time } t\}.
\end{equation*}
The aim of a relational event model is to capture the heterogeneity in the interaction network as well as complex relational and temporal dependencies among events. A model may include all types of predictors commonly used in social network analysis, which can be divided further into three subsets: (i) temporal network effects (such as reciprocity, transitivity, balance), which we will indicate by the variable $s_{ij}(t)$, (ii) fixed network effects such as attributes of the actors, such as gender, age, etc., as well as dyadic covariates such as age difference or social-economic similarity, which we collectively indicate by $x_{ij}$, and (iii) and random network effects, such as (receiver) popularity and (sender) expansiveness, indicated by $z_{ij}$. We define a Cox-type of random effect proportional hazard model for the relational hazard rate,
\begin{eqnarray}
	\label{eq:model}
	\lambda_{ij}(t|x_{ij}(t),z_{ij},s_{ij}(t)=s)&=&\lambda_{s0}(t) \mathrm{e}^{\boldsymbol{\theta}^Tx_{ij}(t)+\boldsymbol{b}^Tz_{ij}},\\
	b &\sim& \mathcal{N}(0,\Sigma(\phi))
\end{eqnarray}
where $\lambda_{s0}(t)$ is a baseline hazard for stratum $s$, $\boldsymbol{\theta}$ are the fixed effects and $\boldsymbol{b}$ is a vector random effects or \emph{frailties}. The frailty terms are assumed to follow a normal distribution with mean zero and a, possibly parametrized, variance matrix $\Sigma(\phi)$. 


\subsection{Network effects}

Inside the random formulation of the relational event model, we identified three different effect types that are driving the interaction dynamics between the actors. In this section we focus on each of these three effect types in more detail.

\paragraph{Endogenous effects.}
When analysing social interactions, one might reasonably expect to see some adherence to social norms, such as reciprocity, triadic closure, or other interaction mechanism such as repetition and assortativity. These mechanisms may increase or decrease the propensity of occurrence of a given action. Reciprocity is a basic characteristic of social life, assuming that individuals tend to establish symmetric patterns of relational events. This dyadic effect describes the flow of exchange between two parties that does not occur simultaneously. 

Triadic closure suggests that the presence of a common third party affects the relation between two individuals. However, there is more than just one way to define a triadic effect in a directed network, see Table \ref{tab:effects}. The cyclic closure describes the relations of generalised exchange, where each individual gives and eventually receives benefits from a different person. Behavioural studies indicate that an individual's cooperative behaviour can be based on prior experiences, regardless of the identity of the other party \citep{rutte2007generalized,fischbacher2001people,isen1987positive}. This mechanism, known as generalized reciprocity \citep{pfeiffer2005evolution} or indirect reciprocity \citep{yarmoshuk2020reciprocity}, assumes that previous receipt of help increases the propensity to help a stranger. Transitive closure describes the process of path shortening, whereby indirect
connections between individuals tend to become direct ties over time. Triadic closure may occur as well through the tie formation arising from similarity in local network position. Sending balance \citep{vu2017relational}, also referred to activity-based structural homophily \citep{robins2009closure}, assumes that two parties may create a tie based on their shared network activity. This effect is analogous to homophily, where the similarity in attributes leads to tie formation. An analogous process of structural homophily is called receiving balance or popularity closure. This effect is based on shared popularity, meaning that individuals may form a connection because they are chosen by the same third party. 

\begin{table}[tb]
	\centering
	\setlength{\tabcolsep}{5mm} 
	\def\arraystretch{1.25} 
	\begin{tabular}{llc}
		\hline\noalign{\smallskip}
		Type & Effect & Structure  \\
		\hline\noalign{\smallskip}
		Dyadic & Reciprocity & 	
%
%
\resizebox{2cm}{!}{\includegraphics{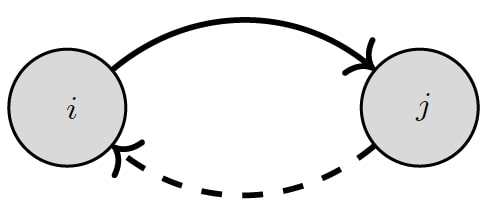}}
		\\
		\hline\noalign{\smallskip}
		\multirow{ 6 }{*}{Triadic closure} & 	Cyclic closure & 
		\resizebox{2cm}{!}{\includegraphics{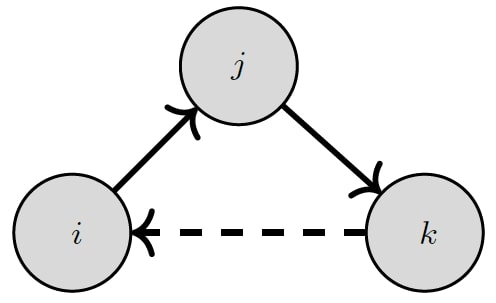}}
%
		\\
		& Transitive closure &
		\resizebox{2cm}{!}{\includegraphics{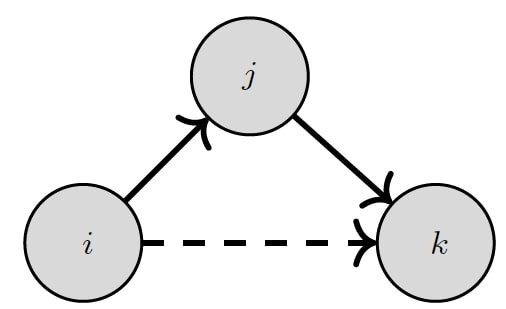}}
%
		\\
		&	Sending balance &
		\resizebox{2cm}{!}{\includegraphics{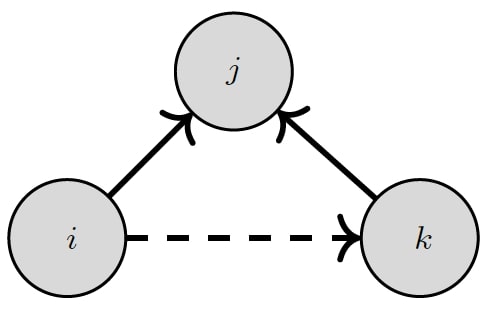}}		
%
		\\
		&	Receiving balance &
		\resizebox{2cm}{!}{\includegraphics{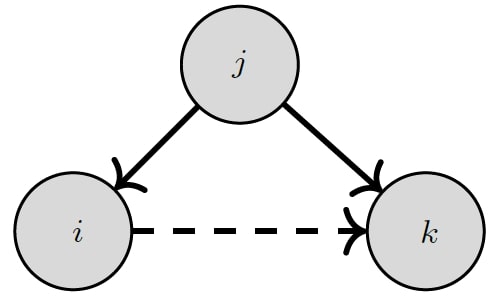}}
%
		\\
		\noalign{\smallskip}\hline
	\end{tabular}
\caption{Common structural network effects for a directed network}
\label{tab:effects}
\end{table}

A special type of endogenous network effects are those associated with measures of nodal centrality reflected in degree-based statistics. A node's emergent expansiveness can be quantified by the number of ties that originate from it, while the number of relational events that are directed towards a node, is an emergent proxy of its popularity. The sender out-degree statistic measures how expansive the current sender has been in the past, i.e., how often they initiated relational events in the past. It is often defined in an exponentially weighted form \citep{lerner2013modeling},  
\begin{equation*}
	\mbox{sender out-degree}(i,t) = \sum_{(i,j,t_e), t_e < t}\mathrm{e}^{-(t-t_e) \frac{\ln 2}{T}}\frac{\ln 2}{T},
\end{equation*}
where the sum is over all past events that included $i$ as sender. The quantity $T$ is a half-life parameter determining at which rate the weights of past events should be reduced. Relational events are weighted in order to give more importance to more recent events. Note that $T=\infty$ corresponds to the unweighted sender out-degree. The sender in-degree measures how often the current sender was
targeted by others in the past, i.e., this measure defines how popular the sender was in the past,
\begin{equation*}
	\text{sender in-degree}(i,t) = \sum_{(j,i,t_e), t_e < t}\mathrm{e}^{-(t-t_e) \frac{\ln 2}{T}}\frac{\ln 2}{T},
\end{equation*}
The receiver out-degree measures how often the current receiver initiated relational events in the past, i.e., it measures how expansive the receiver has been up till now,
\begin{equation*}
	\text{receiver out-degree}(j,t) = \sum_{(j,i,t_e), t_e < t}\mathrm{e}^{-(t-t_e) \frac{\ln 2}{T}}\frac{\ln 2}{T}.
\end{equation*}
The receiver in-degree measures how often the receiver was
targeted by others in the past. It represents the popularity of the receiver,
\begin{equation*}
	\text{receiver in-degree}(j,t) = \sum_{(i,j,t_e), t_e < t}\mathrm{e}^{-(t-t_e) \frac{\ln 2}{T}}\frac{\ln 2}{T}.
\end{equation*}
Another basic mechanism in relational event models is repetition, also known as inertia. Repetition refers to the tendency of past events to be repeated in the future. In particular, this effects represents the accumulated volume of events from actor $i$ to actor $j$ by time $t$:
\begin{equation*}
	\text{repetition}(i,j,t) = \sum_{(i,j,t_e), t_e < t}\mathrm{e}^{-(t-t_e) \frac{\ln 2}{T}}\frac{\ln 2}{T}.
\end{equation*}
Other endogenous network effects may be used to capture various types of participation shifts that play a role in conversational norms \citep{butts,vu2017relational}. For example, a turn-taking effect describes the situation when the receiver takes over the initiative from the current sender. In this scenario, actor $i$ initiates an event towards actor $j$, and subsequently, $j$ initiates an event with an individual other than $i$. To measure this effect, we can use a statistic representing the elapsed time since the last event that satisfy the aforementioned conditions:
\begin{equation*}
	\text{turn-taking}(j,k,t) = \max_{(i,j,t_e),t_e < t, i\neq k}\mathrm{e}^{-(t-t_e) \frac{\ln 2}{T}}\frac{\ln 2}{T}.
\end{equation*}
Another example is turn-continuing, which refers to scenarios where the sender is preserved in multiple relational events. Thus, it involves an event initiated by actor $i$ towards actor $j$, followed by $i$ initiating an event with another individual:
\begin{equation*}
	\text{turn-continuing}(i,k,t) = \max_{(i,j,t_e),t_e < t,j\neq k}\mathrm{e}^{-(t-t_e) \frac{\ln 2}{T}}\frac{\ln 2}{T}.
\end{equation*}

Often, these structural endogenous network effects have been considered as independent variables capturing network patterns influencing event occurrence. A sliding window technique or a weight function have been proposed to account for the temporal aspect of these fundamental endogenous drivers of social interactions. Another approach allowing for time-varying network effects suggests using stratified baseline hazard \citep{JUOZAITIENE2022296}. Stratification avoids making strong assumptions regarding the temporal structure of network formation mechanisms, such as monotonic decay parameters. This approach estimates the strata-specific baseline hazards,
\begin{equation*}
	\hat{\lambda}_{0s}(t) = \frac{\partial }{\partial t}\widehat{\Lambda}_{0s}(t),
\end{equation*} 
where $\widehat{\Lambda}_{0s}(t)$ is a smooth penalised spline estimate of a cumulative baseline hazard.

\paragraph{Exogenous effects.}
The proposed model \eqref{eq:model} also may incorporate covariate effects representing sender and receiver monadic attributes, such as gender, or dyadic relations, such as living in the same neighbourhood or age difference. These covariates represents how exogenous forces shape network formation.

\paragraph{Random effects.} In traditional relational event models, nodes are assumed to be homogeneous, except for the differences captured in available nodal covariates or in the past dynamics of the network. However, this assumption may be insufficient, especially in the context of social networks. The traits governing an individual's sociality and popularity may be complex. For example, some people tend to communicate more actively based upon their personality traits, such as charisma, that is difficult to quantify. The heterogeneity of individuals can be very important to network formation and directly related to the hazard of experiencing an event since more resistant observations remain in the risk set longer \citep{steele2003discrete}. Therefore, the inclusion of nodal random effects enriches the model by accounting for heterogeneity in the nodes of a network, which could not be captured otherwise.

The node specific random effects represent the propensity for individuals to send (expansiveness) and receive (popularity) ties. The expansiveness effect encapsulates all aspects related to an individual's eagerness to initiate events. Similarly, popularity summarises all individual's features that determine their attractiveness as a receiver. Many other random effects can be defined. In fact, random versions of all the above endogenous and exogenous variables can be considered.

\subsection{Frailty model estimation}

Following \citet{survival-book}, an integrated partial likelihood for the model \eqref{eq:model} is given as
\begin{equation*}
	IPL(\theta,\phi) = \frac{1}{(2 \pi)^{q/2}|\Sigma(\phi)|^{1/2}}\int PL(\theta,b) \mathrm{e}^{-b^T\Sigma^{-1}b/2}\mathrm{d}b,
\end{equation*}
where $q$ is the number of random effects and $PL(\theta,b) = \prod_{o=1}^n \lambda_{i_oj_o}(t_0)/\sum_{(s,r)\in\mathcal{R}(t_o)} \lambda_{sr}(t_0)$ is Cox partial likelihood for any fixed values of $\theta$ and $b$. However, the $IPL$ is an intractable multidimensional integral and to perform computations involving this likelihood we use the Laplace approximation. In this case, the log penalised partial likelihood ($LPPL$) is replaced with a second order Taylor series about its value at the maximum of the function
\begin{eqnarray*}
	LPPL(\theta,b,\phi) &=& LPL(\theta,b) - \frac{1}{2}b^TA^{-1}(\phi)b \\ &\approx& LPPL(\hat{\theta}(\phi),\hat{b}(\phi)) - \frac{1}{2} (b - \hat{b})^TH_{bb}(b - \hat{b}),
\end{eqnarray*}
where $H$ is the matrix of second derivatives of the $LPPL$ and $H_{bb}$ is the portion of this matrix
corresponding to the random effects. When $\phi$ is fixed, the relevant values of $\theta$ and $b$ that maximize the $LPPL$ can be computed using the same methods as a usual Cox regression model. 

\subsection{Likelihood-ratio test}

We can formally test the significance of the random effects using a likelihood-ratio test, which compares the goodness of fit of two nested statistical models. The likelihood-ratio test statistic is defined as follows:
\begin{equation*}
	LR = -2(\ln L_0 - \ln L_1),
\end{equation*}
where $L_0$ and $L_1$ are the maximum likelihood values for the reduced and full models, respectively. This statistic has an asymptotic $\chi^2$ distribution with degrees of freedom equal to the difference in the number of parameters between the two models. We propose to perform the likelihood-ratio test based on the approximate integrated partial likelihood. 

\section{Simulation studies}

In order to test the performance of the proposed framework and assess the consequences of neglecting nodal heterogeneity, we simulate and examine networks in four sets of analyses. The first two simulation studies analyse whether the proposed frailty approach recovers the true parameters, and subsequently how its estimates of the model parameters improve under increasing sample size scenarios. In the third simulation study we show how nodal heterogeneity induces ghost triadic effect, even when including traditional nodal degree statistics in the model. The objective of the fourth simulation study is to illustrate the usefulness of the partial likelihood-ratio test for the inclusion of the random effects. 

\subsection{Network effects recovery}
\label{subsec:sim1}

A simulation-based experiment is conducted to demonstrate that the proposed approach is able to adequately recover the underlying parameters. To analyse the relative bias in parameters estimated with correctly specified frailty models, we consider the following simulation scenario. In each of 20 replications, we simulate 10,000 events among 100 individuals. The rate of each event is defined following the assumption that triadic closure effects have no impact on link occurrence. The baseline hazard function is set to be a constant equal to 1, assuming that waiting times are exponentially distributed. Therefore, the rate of each event depends only on the nodal random effects. The popularity and expansiveness random effects are generated from a normal distribution, i.e., $b_i^{pop} \sim \mathcal{N}(0,1.3^2), b_i^{exp} \sim \mathcal{N}(0,0.9^2)$.
Each dataset is analysed by the relational event model with frailty given in \eqref{eq:model}. Four stratified models are fitted to the generated datasets focusing on one triadic closure effect at a time.

\begin{figure}[tb]
	\centering
	\begin{tabular}{cc}
		\includegraphics[width=0.5\linewidth]{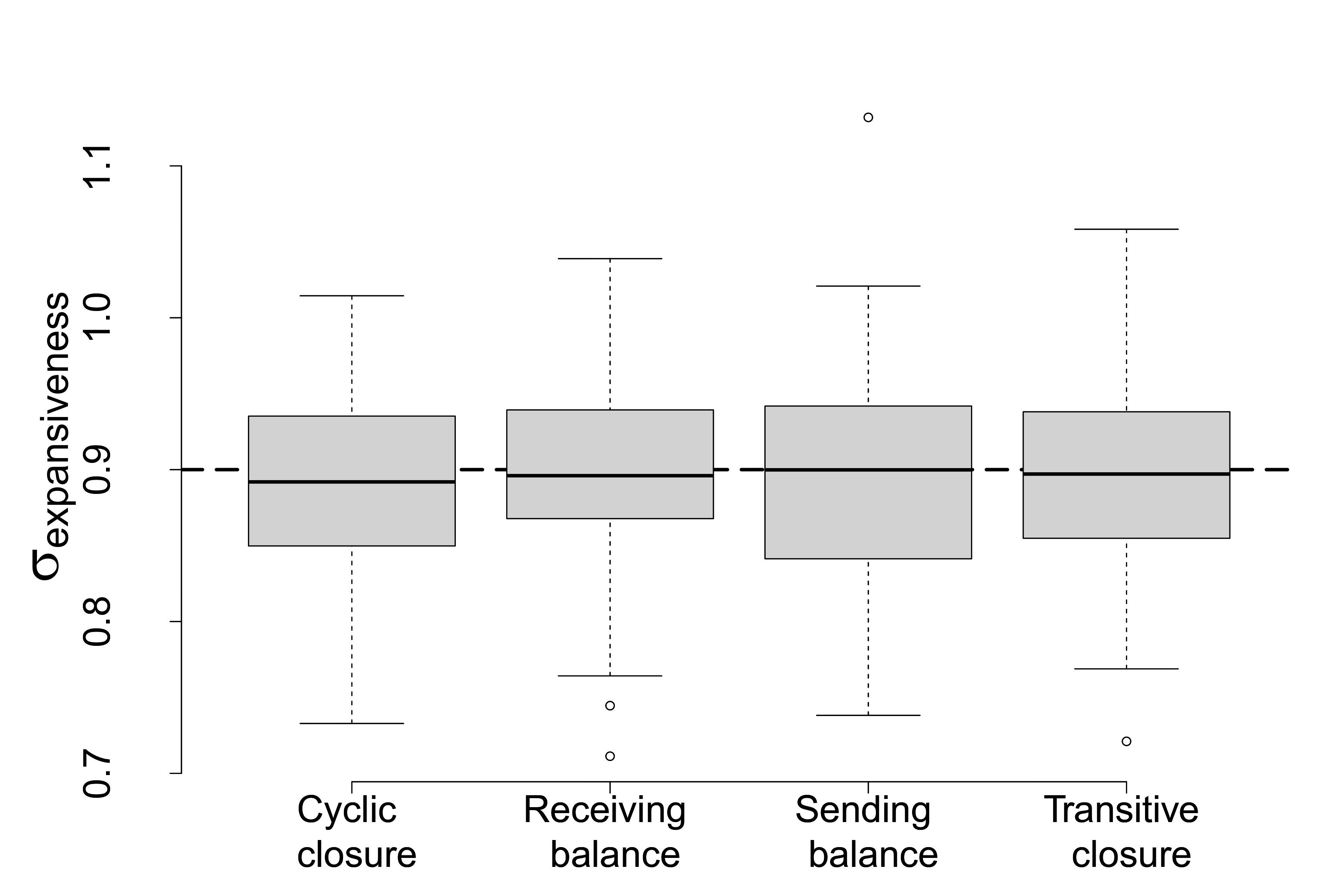}
		&
		\includegraphics[width=0.5\linewidth]{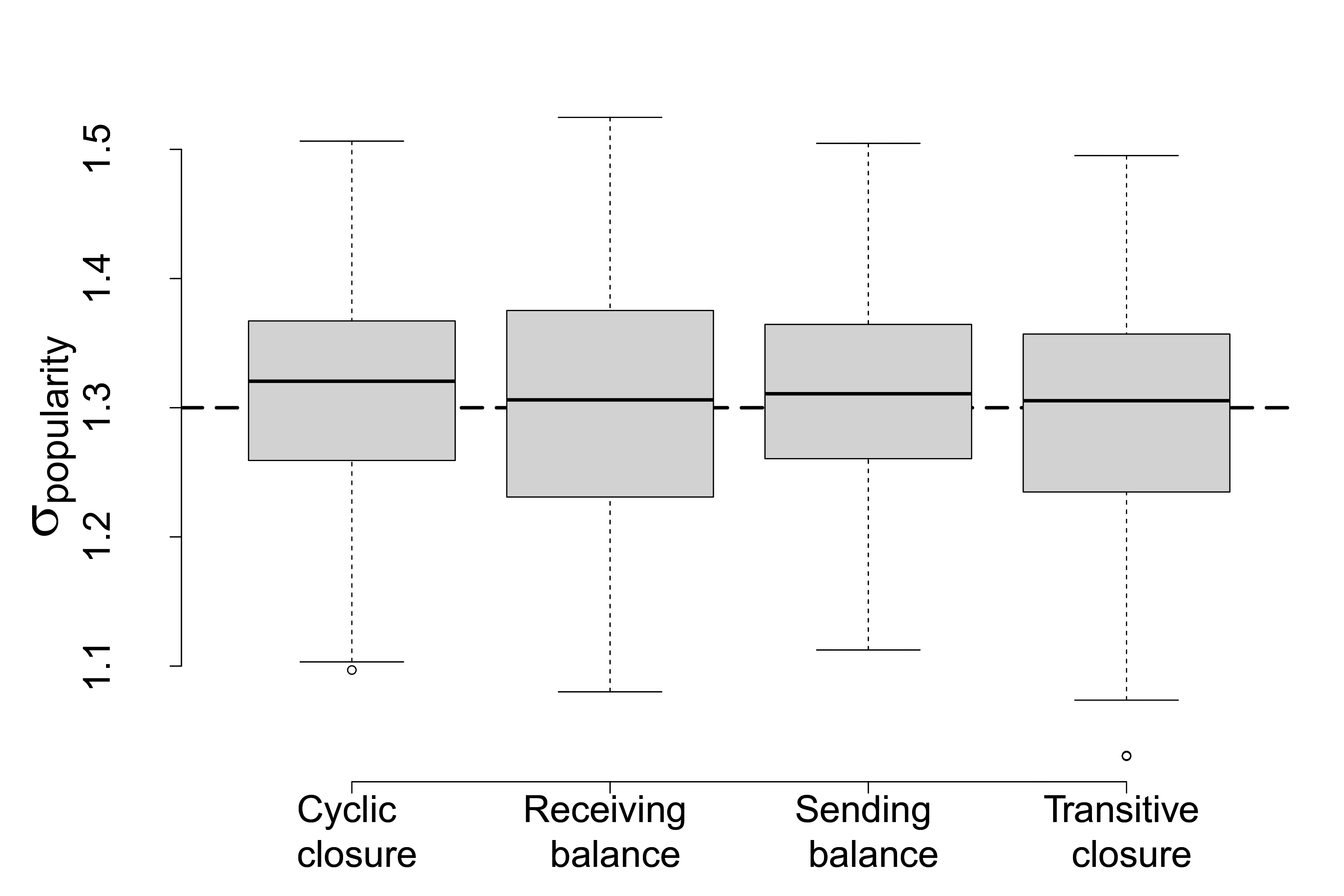}
		\\ (a) & (b)
	\end{tabular}
	\caption{Boxplot of the estimated standard deviations of (a) expansiveness random effect (0.9) and (b) popularity random effect (1.3) in four different model formulations.}
	\label{fig:senderreceiver}
\end{figure}

\begin{figure}[tb]
	\centering
	\includegraphics[width=0.7\linewidth]{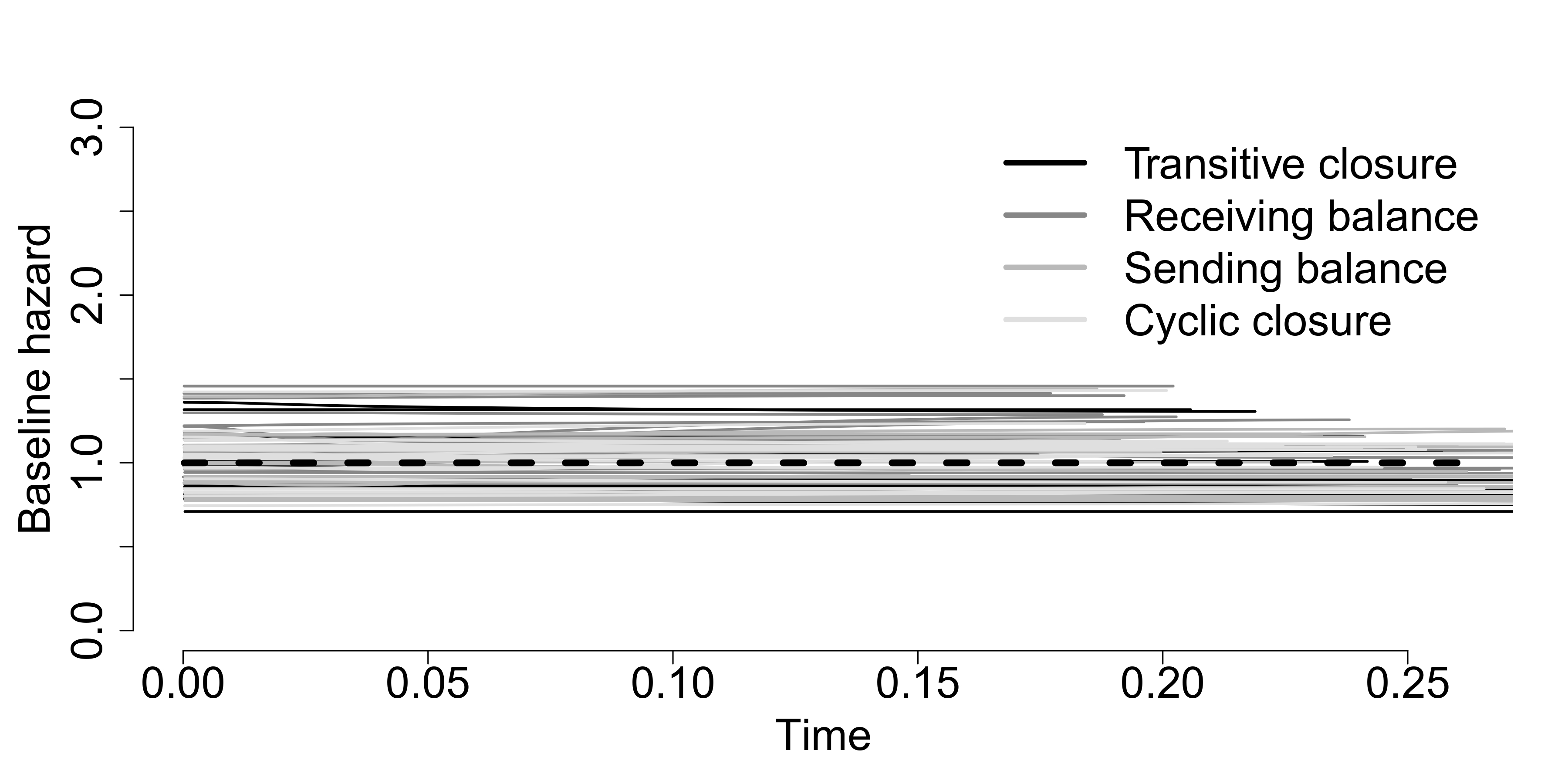}
	\caption{True (dashed) and estimated (solid) baseline hazard curves: different colours represent the results of four random effect models focusing on different triadic closure effects. There is no obvious bias for any of the models.}
	\label{fig:baseline_h}
\end{figure}

The standard deviations of the random effects are recovered to a fairly high level of accuracy (see Figure \ref{fig:senderreceiver}). The proposed frailty model is able to estimate the random effect standard deviation values with no noticeable bias patterns, regardless of the triadic closure model. That is, both standard deviation estimates are centred on their true values (0.9 and 1.3, respectively). Moreover, the estimated smooth baseline hazard curves in Figure \ref{fig:baseline_h} indicate that the frailty approach recovers the underlying process well. The proposed framework adequately recovers the shape of the underlying distribution with no obvious bias in its magnitude. 

\subsection{Effect of increasing sample size}

To assess the effect of varying sample size on estimator performance, we vary two factors: number of individuals (i.e., the number of random effects levels) and the number of events (number of observations). For each scenario, we generate data under previously presented settings.  Nodal heterogeneity is introduced in the form of the random effects that are drawn from a normal distribution with $\sigma_{pop} = 0.9, \sigma_{exp}=0.5$. To explore the importance of varying number of events we simulate three datasets under different scenarios, i.e., we simulate 500, 1,500 and 4,500 events among 100 individuals. Accordingly, to analyse the effect of varying number of individuals we simulate 10,000 events among 10, 30 and 90 individuals. Each dataset is analysed using the previously described frailty model. 

\begin{figure}[tb]
	\begin{tabular}{cc}
		\includegraphics[width=0.45\linewidth]{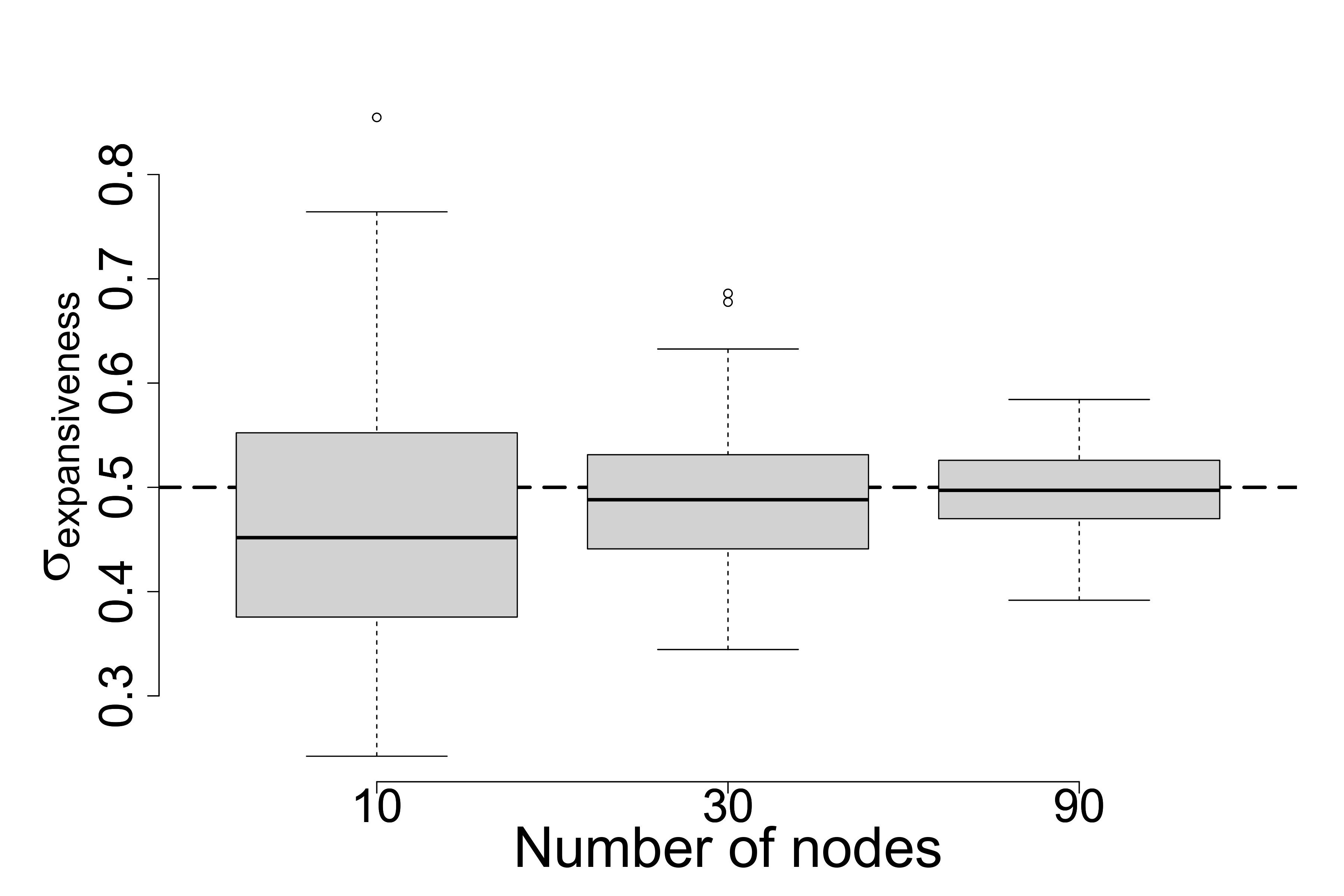}
		&
		\includegraphics[width=0.45\linewidth]{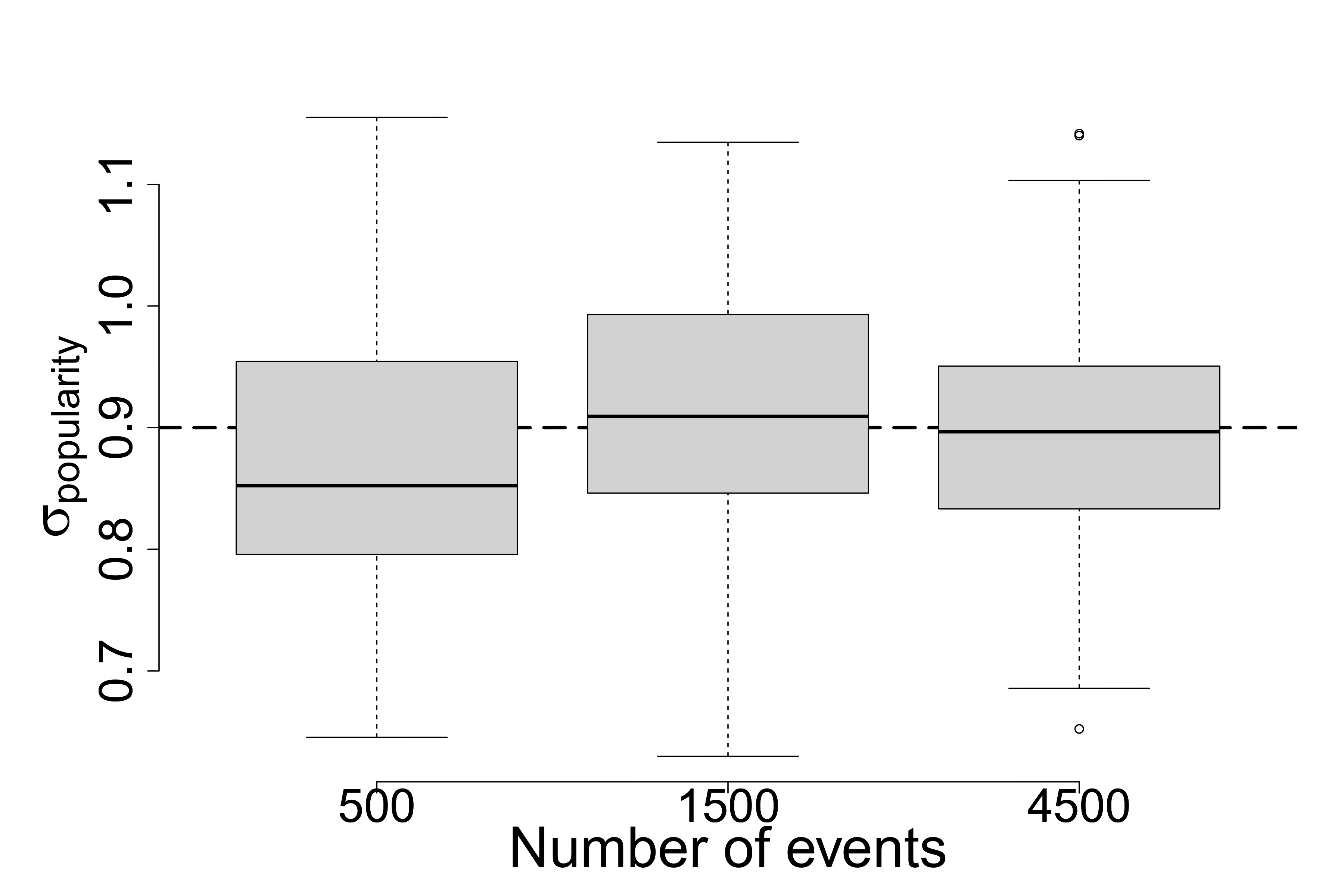} \\
		(a) Expansiveness vs nodes
		&
		(b) Popularity vs edges
	\end{tabular}
	\caption{(a) typical $1/\sqrt{n}$ effect of network size on the estimate of the expansiveness standard deviation; (b) almost no effect of number of relational events on the estimate of the popularity standard deviation.}
	\label{fig:nodes-edges}
\end{figure}

Figures \ref{fig:nodes-edges} summarises the results of 100 replications. Figure \ref{fig:nodes-edges}a demonstrates the effects of increasing the number of individuals on estimator performance by providing the estimated standard deviations for three different scenarios. The results for the popularity standard deviation are similar. As expected, the uncertainty around standard deviation estimates generally decreases with an increasing number of random effects levels in an approximate $1/\sqrt{n}$ fashion (here: proportional to $1/\sqrt{10}$, $1/\sqrt{30}$ and $1/\sqrt{90}$, respectively). Results presented in Figure \ref{fig:nodes-edges}b indicate that increasing the number of relational events has only a minor effect on random effect estimation accuracy. The reason is that the main limiting factor is the number of simulated random effects, which remains constant (here: 100).

\subsection{Expansiveness and popularity induce ghost triadic effects}

Another simulation study is conducted to demonstrate the consequences of not accounting for nodal random effects in relational event models. In particular, we are interested in how heterogeneity affects the estimates of triadic closure effects. We consider the same simulation scenario as above, 
including only random node effects for popularity and expansiveness. We fit two models. The first includes only endogenous effects, such as reciprocity and triadic closure, whereas the second also includes traditional nodal degree statistics to account for nodal heterogeneity. Both models ignore the random effects, as in a traditional relational event model.


\begin{figure}[tb]
	\begin{tabular}{cc}
		\includegraphics[width=0.5\linewidth]{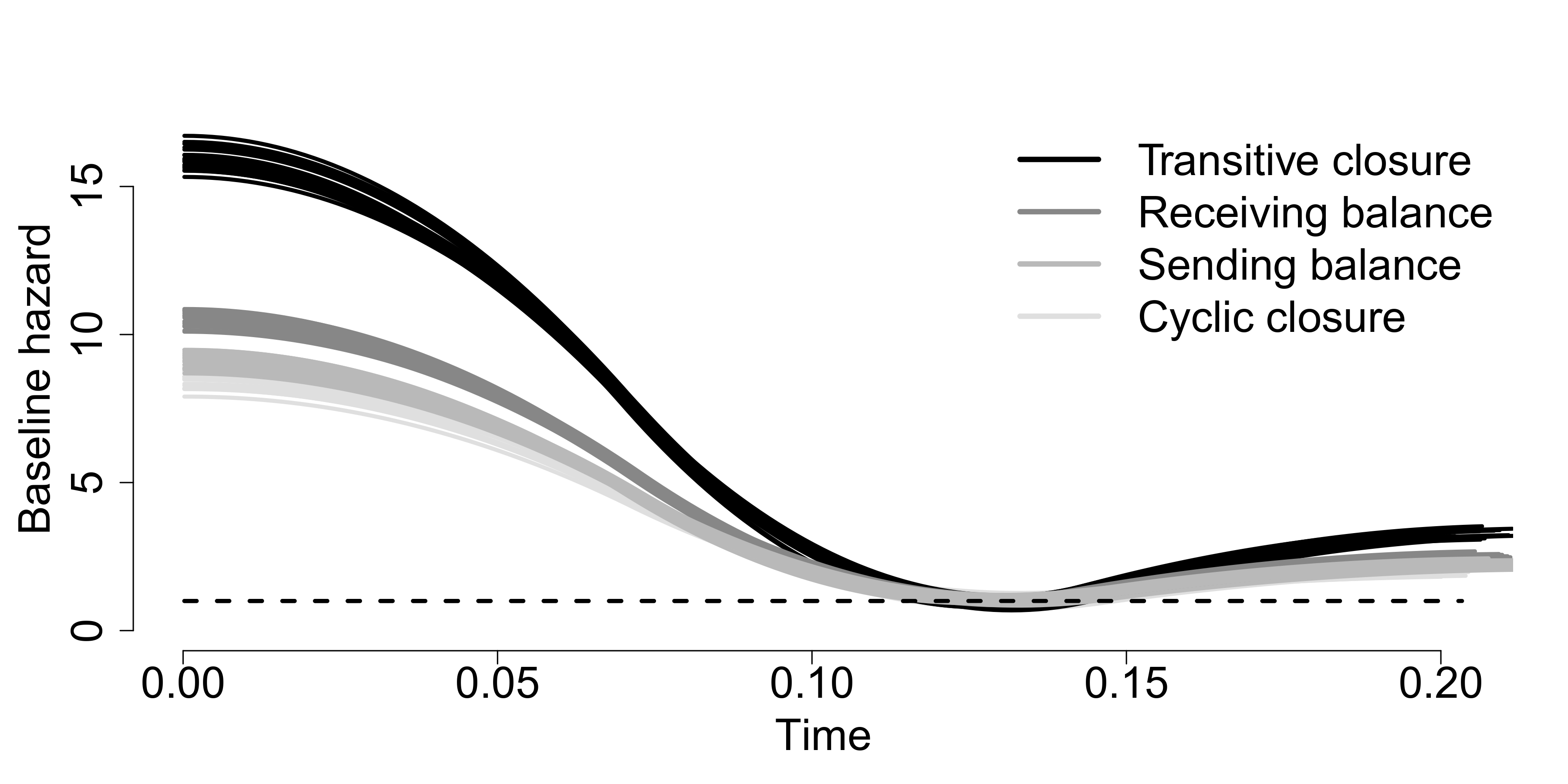}
		&
		\includegraphics[width=0.5\linewidth]{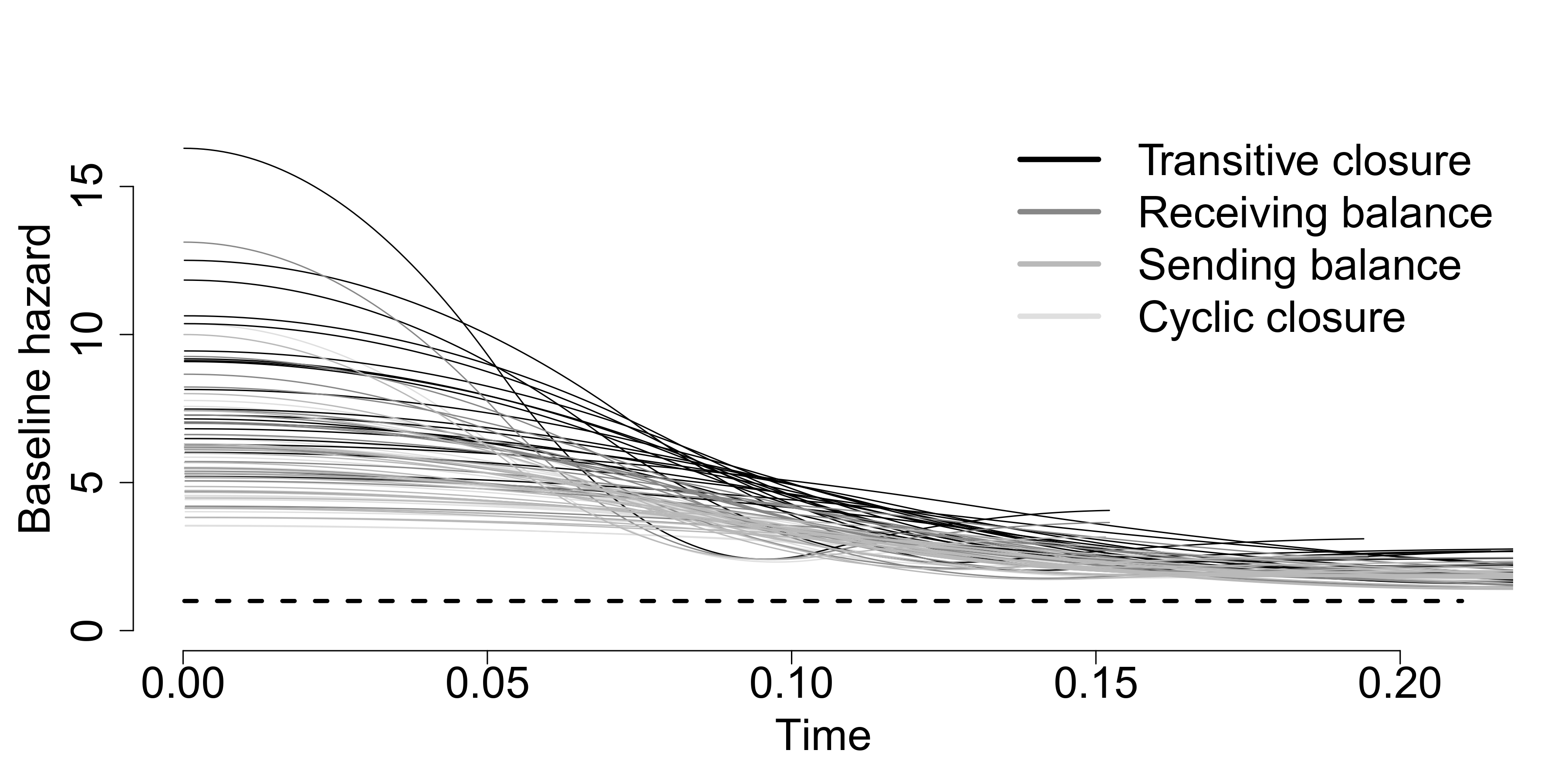} \\
		(a) & (b)
	\end{tabular}
	\caption{(a) Phantom triadic effects are identified as a result of the presence of random popularity and expansiveness effects. (b) Degree- and intensity-based statistics are unable to fully account for nodal heterogeneity. In both plots the solid curves depict the overestimated triadic closure effects, while the dashed line represents the actual level of triadic closure.}
	\label{fig:triadic}
\end{figure}

In the first simulation study we only fit four triadic effects based on a stratified relational event model, that also includes reciprocity. Figure~\ref{fig:triadic}(a) shows solid lines of the estimated hazard functions for each triadic closure effect. The dashed line is the true baseline hazard function. It is immediately apparent that the estimates from the stratified relational event model severely overstate the triadic effects, particularly in the beginning. These results confirm that failure to account for nodal heterogeneity can vastly overestimate the actual amount of triadic closure present in the network. Furthermore, it is interesting to note the different bias pattern among the four triadic closure effects. The most severe bias is observed for the transitive closure effect. This suggests that transitive closure is the most sensitive to heterogeneity, and it responds to misspecification by significantly increasing in magnitude. Intuitively, this is due to the nature of transitive closure. This structural effect implies a local hierarchy, with one node only receiving ties, one node only sending ties and one neutral node receiving and sending ties. Therefore, the receiving node is the most popular within the group, and the sending node is the most expansive within it. For this reason, the effect of shared partners might be overestimated in order to compensate for underestimating the effects of social behaviour. On the other hand, in a cyclic closure, the direction of all ties is consistent and none of the three nodes would be singled out, which corresponds with the simulation study results indicating that cyclic closure is more robust to nodal heterogeneity.

\begin{figure}[tb]
	\centering
	\includegraphics[width=\linewidth]{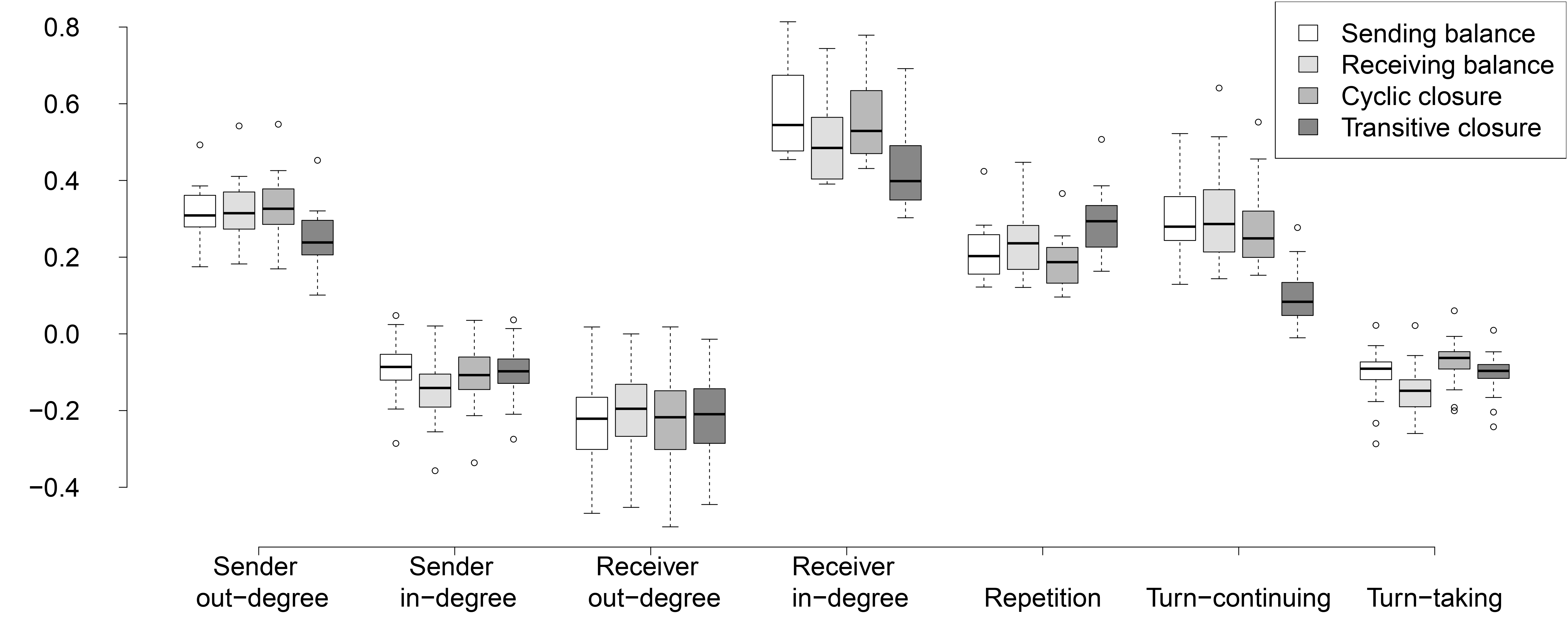}
	\caption{For each triadic closure effect, the boxplot represents the distribution of the parameter estimates obtained over 20 replications. In all cases, we observe positive receiver in-degree and sender out-degree effects.}
	\label{fig:est}
\end{figure}

It can be argued that our first analysis is too simplistic. The usual approaches to account for nodal heterogeneity suggest including degree- and intensity-based network effects. Therefore, for each triadic closure effect, we also estimate the model, incorporating sender and receiver in-degree and out-degree, repetition, turn-taking and turn-continuing effects. The distributions of the estimated effect sizes are summarized in Fig. \ref{fig:est}. Notably, we observe positive estimates for sender out-degree and receiver in-degree effects, indicating that these network statistics can capture a portion of individuals' popularity and expansiveness. However, the included degree- and intensity-based network effects do not entirely account for variations between individuals. Figure~\ref{fig:triadic}(b) shows that the estimated baseline hazard functions for each triadic effect still tend to overestimate the actual amount of triadic closure. Thus, we can conclude that degree- and intensity-based statistics can be used to reduce nodal heterogeneity to some extent. However, they are insufficient in fully accounting for nodal heterogeneity and obtaining credible estimates of other endogenous network effects, such as transitive closure.

\subsection{Performance of partial likelihood-ratio test}

The previous section demonstrates the severe consequences of a failure to account for nodal heterogeneity that is not explained by exogenous covariates. In fact, there is greater harm in excluding a random effect that is necessary than including a random effect that is not needed \citep{gelman2006data}. Nevertheless, from the point of view of parsimony and interpretation, we want a tool that is able to test for the need to include random effects. In this section, we test the partial likelihood-ratio test based on the integrated partial likelihood.

\begin{figure}[tb]
	\centering
	\includegraphics[width=0.5\linewidth]{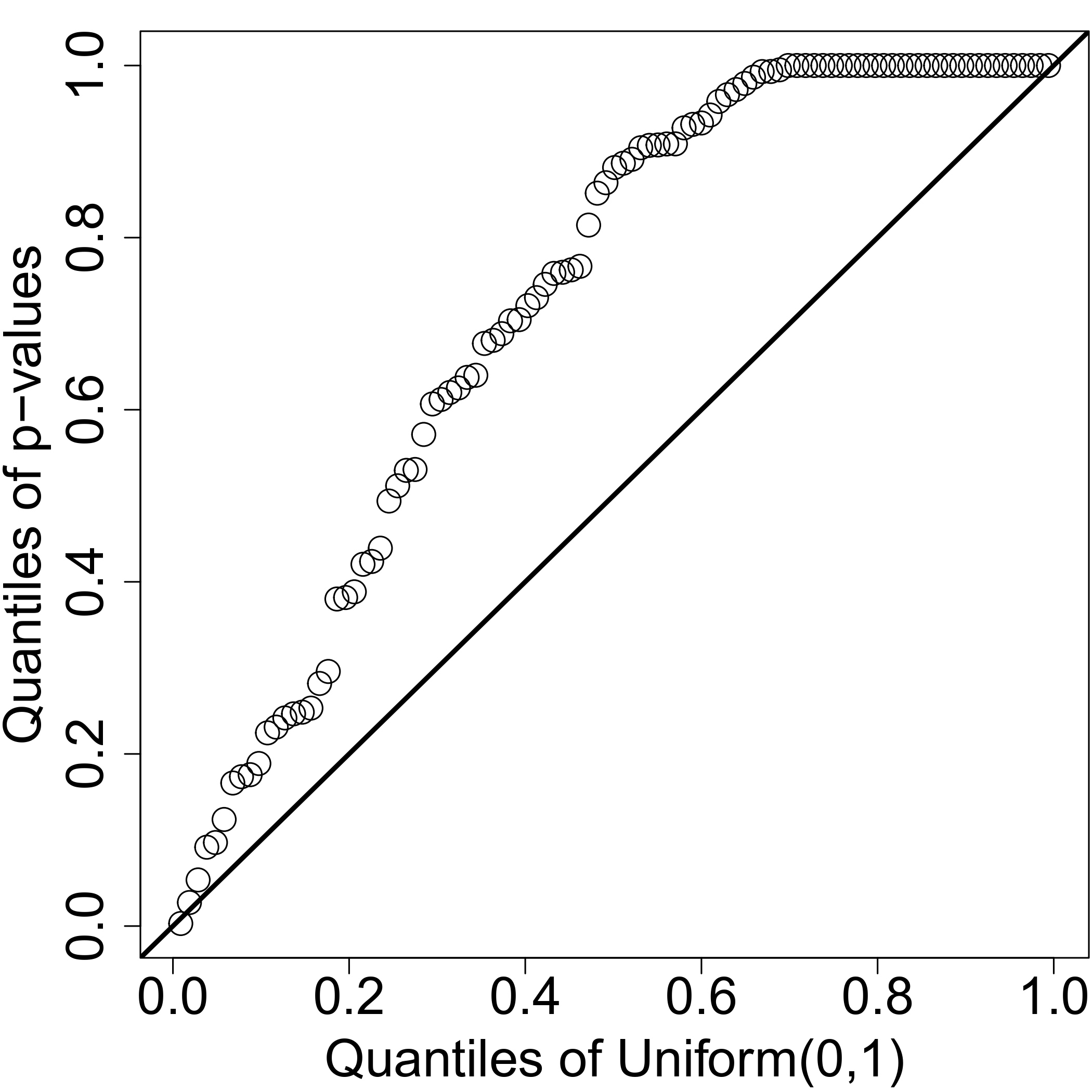} 
	\caption{QQplot comparing the uniform distribution to the $p$-values for partial likelihood ratio test, suggesting that the likelihood ratio test based on the approximate integrated partial likelihood is slightly conservative.}
	\label{fig:qqplot}
\end{figure}

We simulate 10,000 events among 100 individuals assuming that waiting times are exponentially distributed. Spontaneous event times follow an exponential distribution with parameter $\lambda_{sp} = 1$, reciprocal events have a higher risk $\lambda_{re} = 2$, parameters for transitive and R+T strata events are respectively equal to $\lambda_{tr} =3$ and $\lambda_{r+t}=4$. Thus, we assume that network dynamics is driven by the triadic effects, and there are no random effects. 

For each dataset, we fit two stratified relational event models models with and without frailty terms. Fitted models are compared using integrated version of the partial likelihood-ratio test. Figure \ref{fig:qqplot} shows the QQplot comparing the quantiles of the calculated $p$-values to the quantiles of the uniform distribution. We can see that $p$-values are slightly conservative. This means that (i) this test tends not to include random effects, when they are not needed, and (ii) it requires a bit more evidence, when they are needed.

\section{Illustrative case studies}

While the previous section demonstrated the potential importance of directly modelling nodal heterogeneity in simulated network data, there remains a question as how useful the random popularity and expansiveness model is in practical, real-world problems. For this reason, in this section we analyse six real world datasets as illustrative examples of the importance of accounting for nodal heterogeneity. 
\begin{enumerate}
	\item Manufacturing company \citep{manufacturing_email}: a dynamic network describing the internal email communication between employees of a mid-sized manufacturing company. This study contains 82,614 email communications observed among 176 employees over nine months period beginning in January 2010.
	\item Enron email \citep{Klimt2004TheEC}: 2934 emails among 119 individuals between July 2001 and August 2001. 
	\item Classroom \citep{Mcfarland}: 691 communication events recorded between 20 individuals within a high school classroom. Data also contains two actor-level covariates defining the individual's gender and role (i.e. student or teacher).
	\item Phone calls \citep{sapiezynski2019interaction}: 3,600 phone calls among 540 students observed over a period of four weeks. The dataset was collected via smartphones as part of the Copenhagen Networks Study.
	\item Social evolution \citep{madan2011sensing}: 439 phone calls observed among the 54 students residing in a university dormitory. The dataset also includes two exogenous variables: the floor of the dormitory on which the student resides, and the grade type of each student (i.e. freshmen, sophomore, junior, senior, or graduate tutors). 
	\item Virtual Battlespace 2 (VBS2) game \citep{pilny2016illustration}: 299 communications among 4 players who were engaged in a VBS game scenario. 
\end{enumerate}

We have slightly preprocessed datasets by removing instances when sender and receiver coincide, as well as events occurring simultaneously but having different senders, as such events might create situations where an open triad is created and closed at the same time. We did allow for multicast events, where communication events are sent to multiple receivers. In these applications, multiple events occurring at the exact same time are treated as ties. In order to deal with tied events, we use Efron's approximation because it is accurate and computationally efficient \citep{survival-book}. 

To analyse how nodal heterogeneity affects the estimates of triadic closure effects for each dataset we fit the stratified relational event frailty model with sender and receiver random effects. Both models include a list of endogenous network effects, i.e., in-degree and out-degree statistics for sending and receiving nodes, repetition, turn-taking, turn-continuing, reciprocity, triadic effects. For the email communication datasets (Manufacturing company and Enron email) we also included an off-set $\ln(n_{r(e)})$, the natural logarithm of the number of receivers per email $n_{r(e)}$.

Based on the AIC, we identify the most appropriate triadic effect for each dataset. We find that the transitive closure effect is the most suitable for the Manufacturing company data, while the sending balance effect is the most appropriate for Phone calls, Enron email and Social evolution datasets. For the Classroom data we use the cyclic closure effect. Only, the VBS2 game dataset does not exhibit any triadic closure effect resulting in the model including only the time-varying reciprocity effect. These estimates are shown in Table~\ref{tab:rec-tri}.

	\begin{longtable}{lcc}
		\toprule
		& Reciprocity & Triadic effect  \\
		\midrule
		\endhead
		\bottomrule
		\multicolumn{3}{c}{\textsl{Continues on the next page}}
		\endfoot
		\hline
		\caption{Estimated reciprocity and triadic effects for fixed and mixed effects models, showing that unmodelled nodal heterogeneity severely distorts these effects in the Manufacturing company, Classroom and Phone calls datasets.}\\
		\label{tab:rec-tri}
		\endlastfoot			
		Manufacturing & \adjustbox{valign=c}{\includegraphics[width=0.3\linewidth]{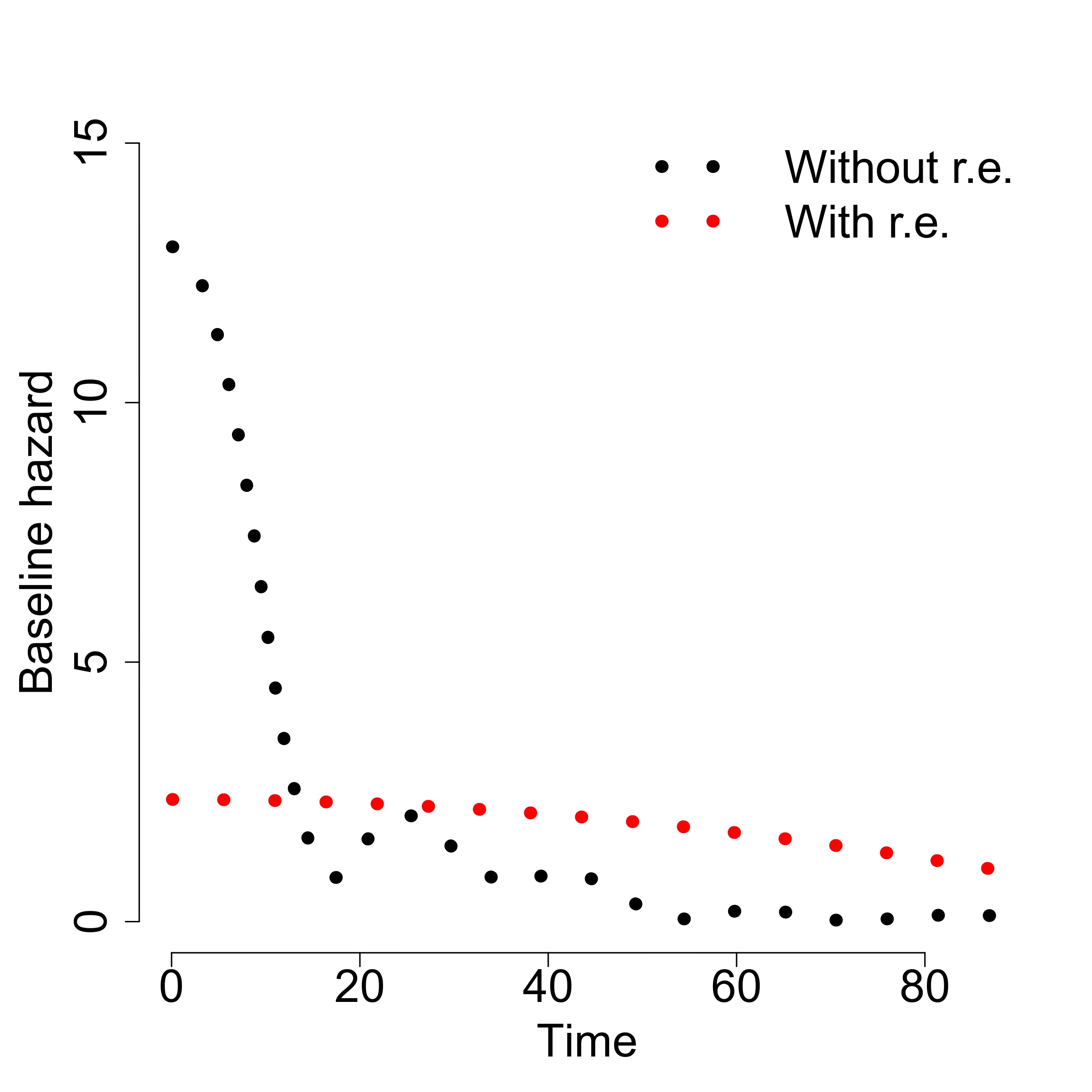}} & \adjustbox{valign=c}{\includegraphics[width=0.3\linewidth]{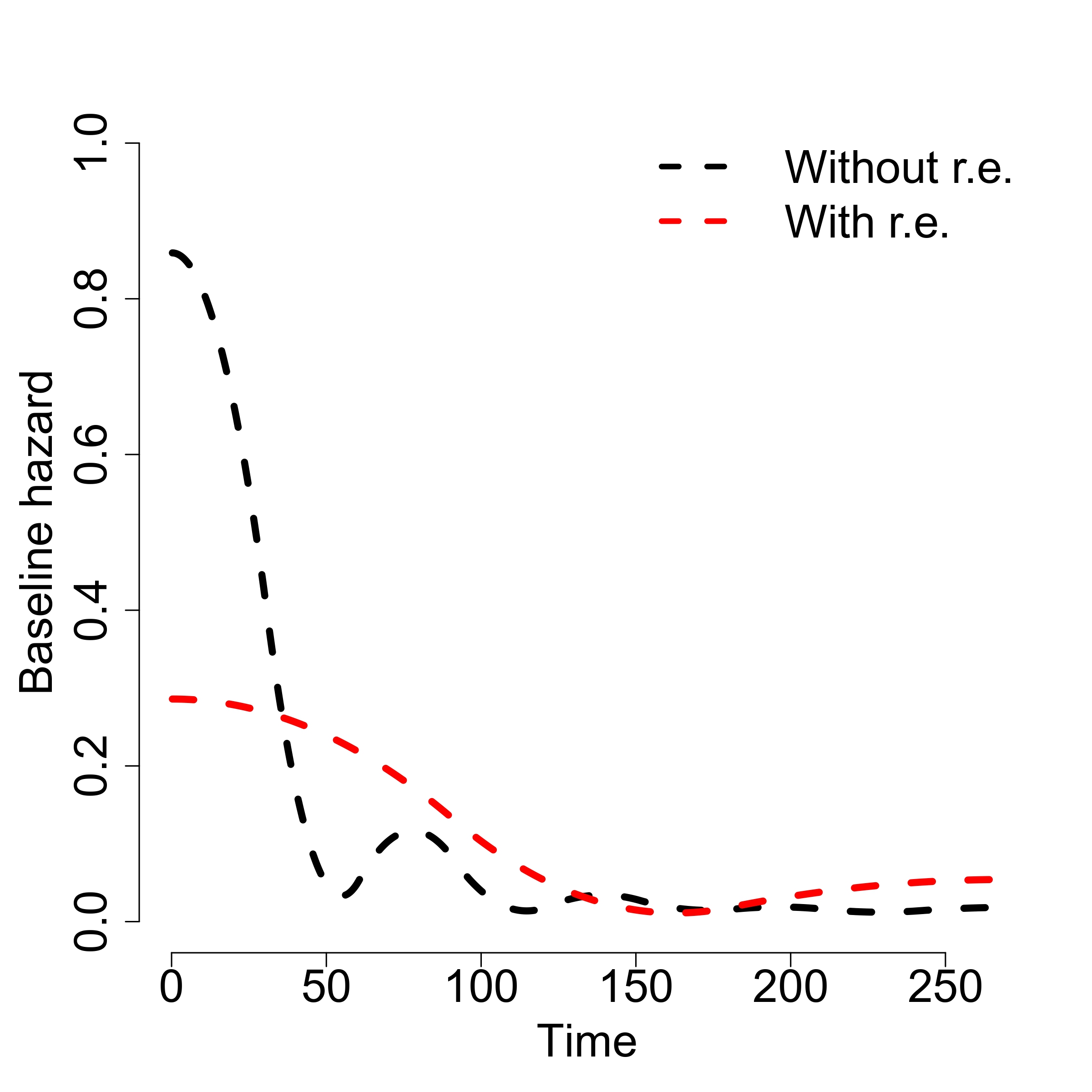}}\\
		\midrule
		Enron email & \adjustbox{valign=c}{\includegraphics[width=0.3\linewidth]{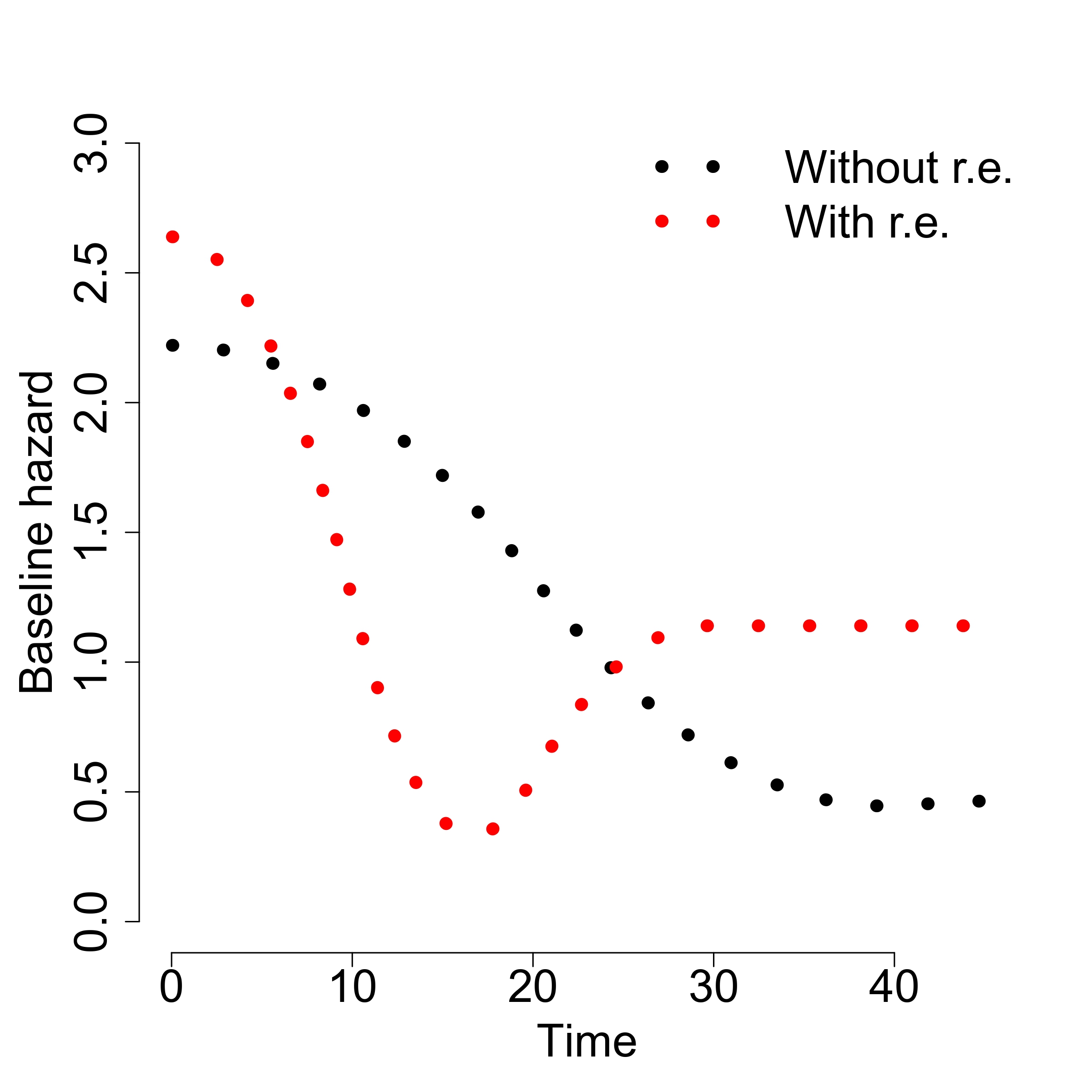}} & \adjustbox{valign=c}{\includegraphics[width=0.3\linewidth]{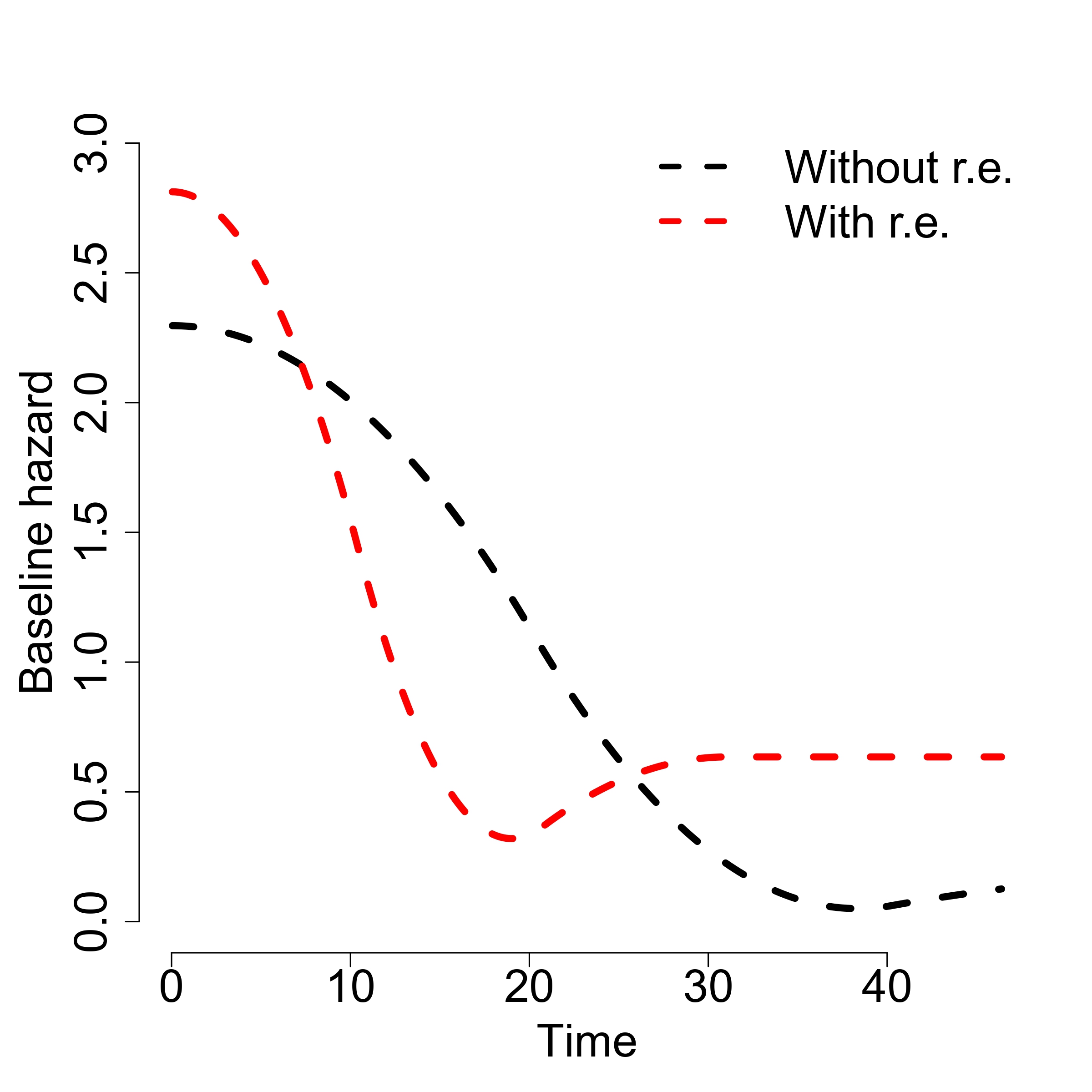}}\\
		\midrule
		Classroom & \adjustbox{valign=c}{\includegraphics[width=0.3\linewidth]{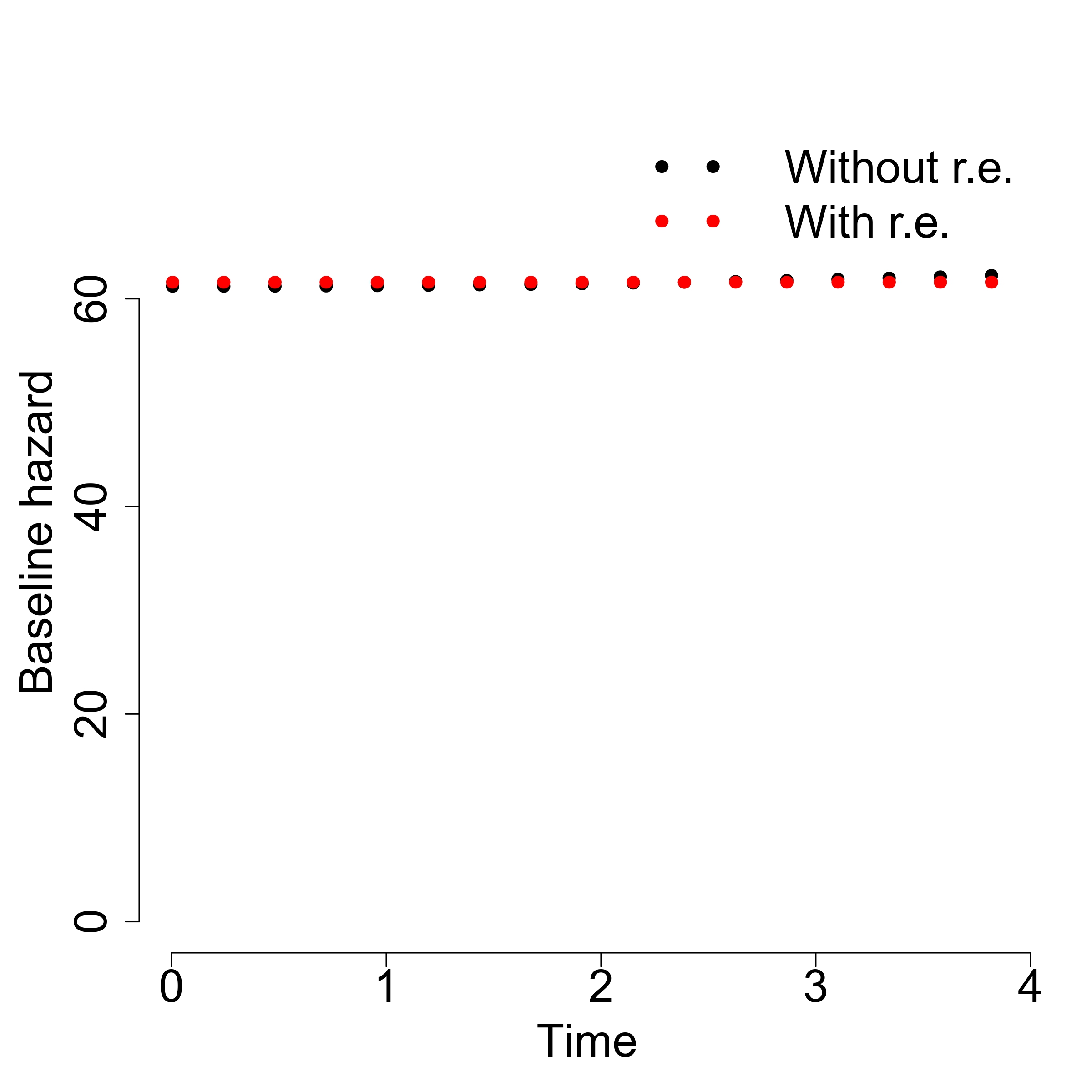}} & \adjustbox{valign=c}{\includegraphics[width=0.3\linewidth]{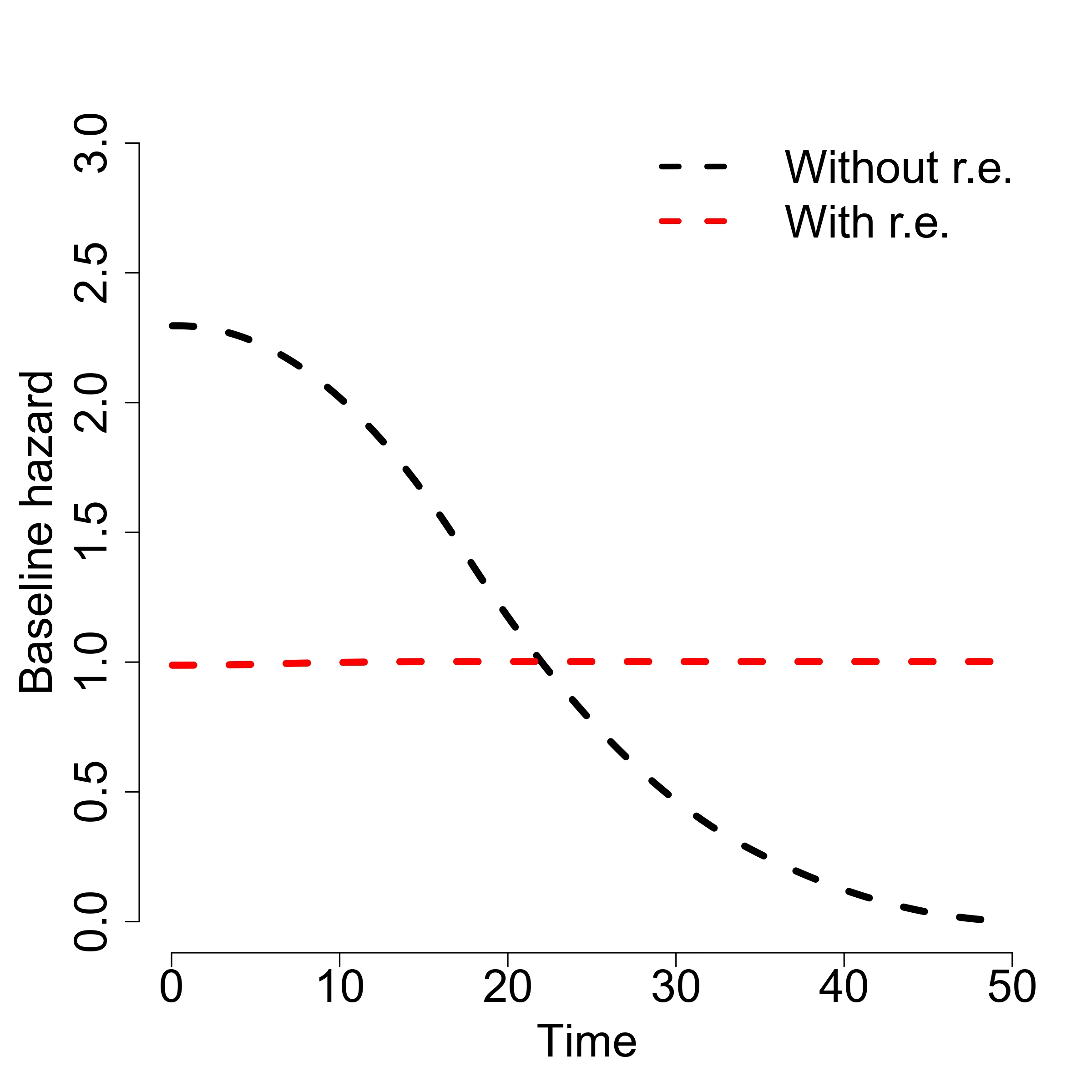}}\\
		Phone calls &  
		\adjustbox{valign=c}{\includegraphics[width=0.3\linewidth]{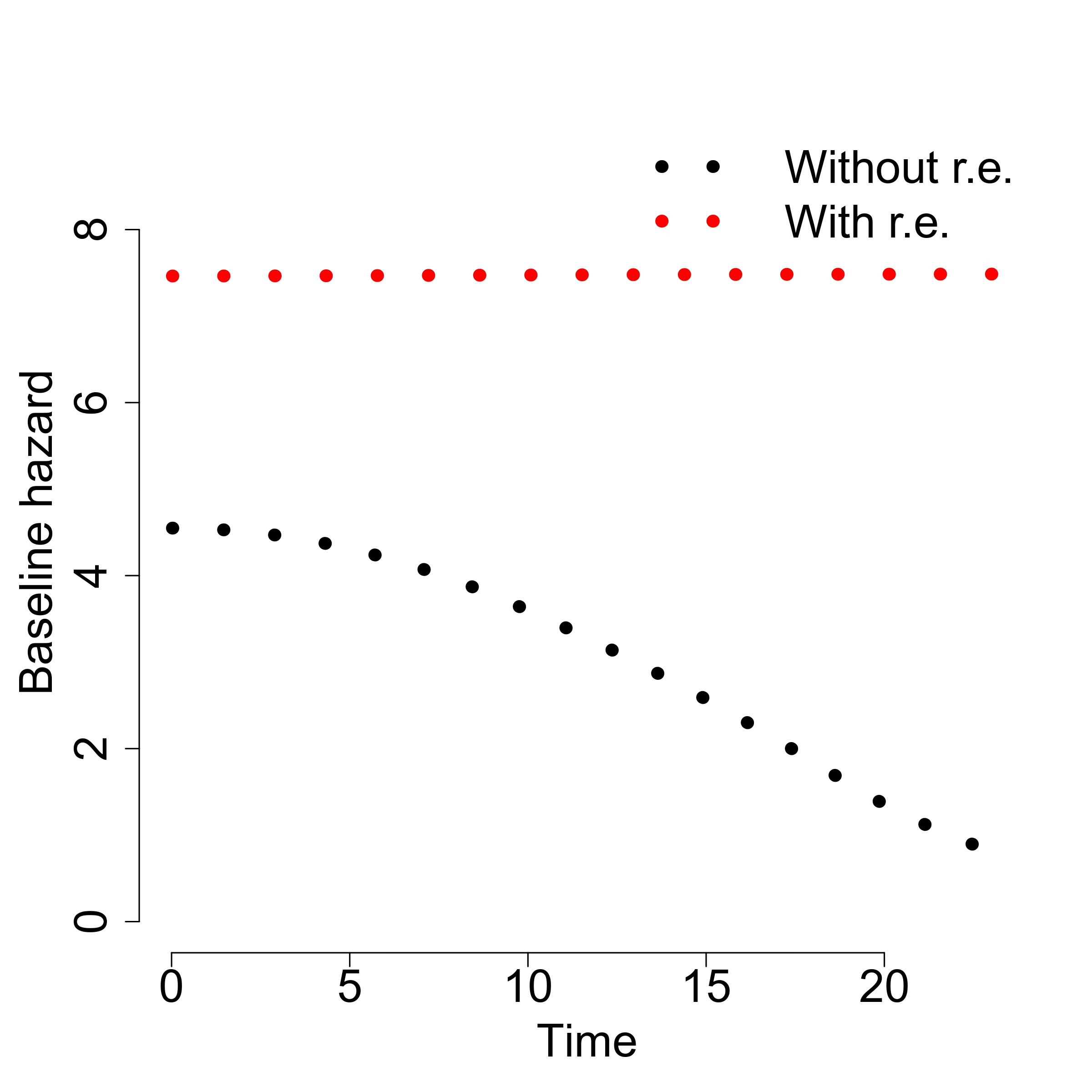}} & \adjustbox{valign=c}{\includegraphics[width=0.3\linewidth]{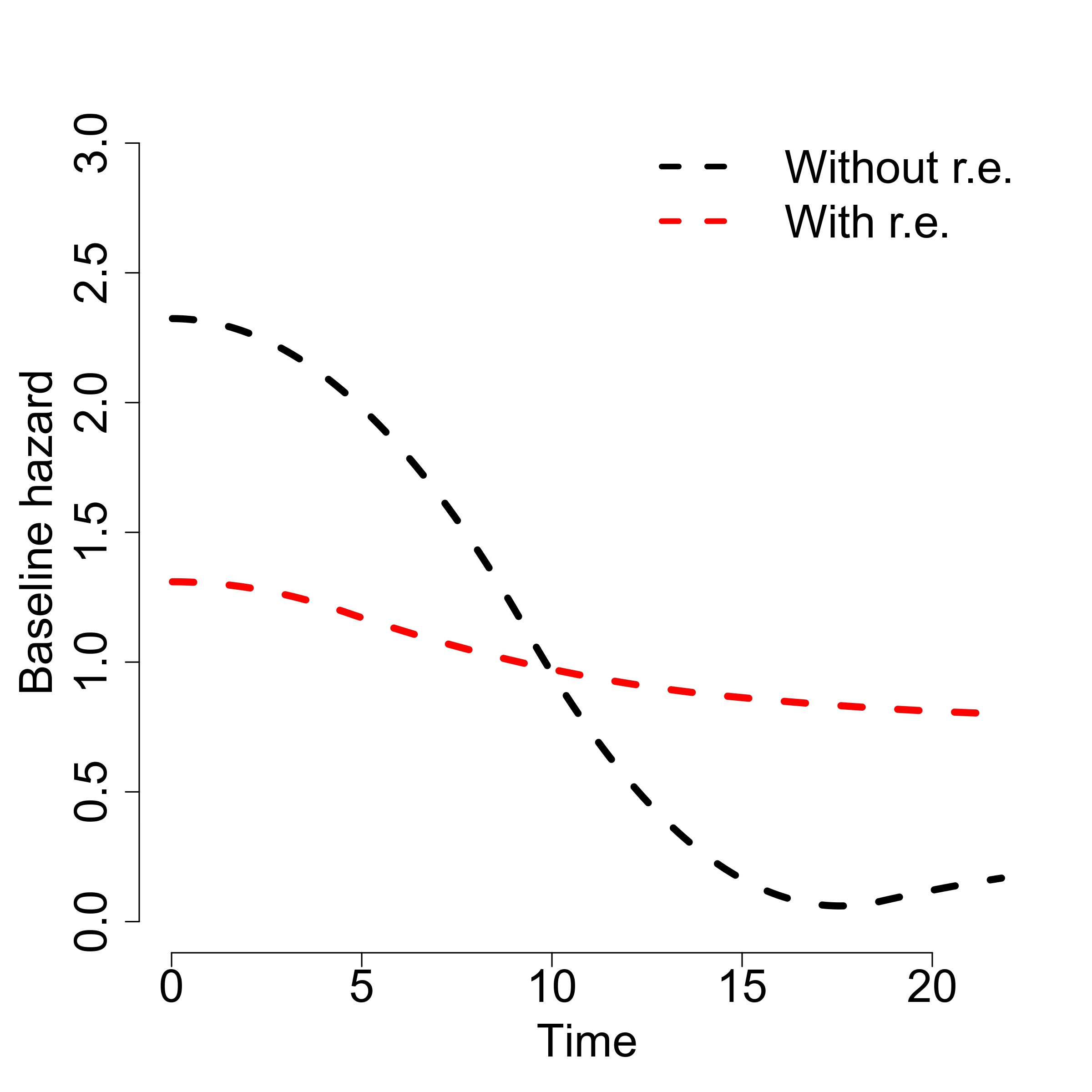}}\\
	\midrule
		Social evolution & \adjustbox{valign=c}{\includegraphics[width=0.3\linewidth]{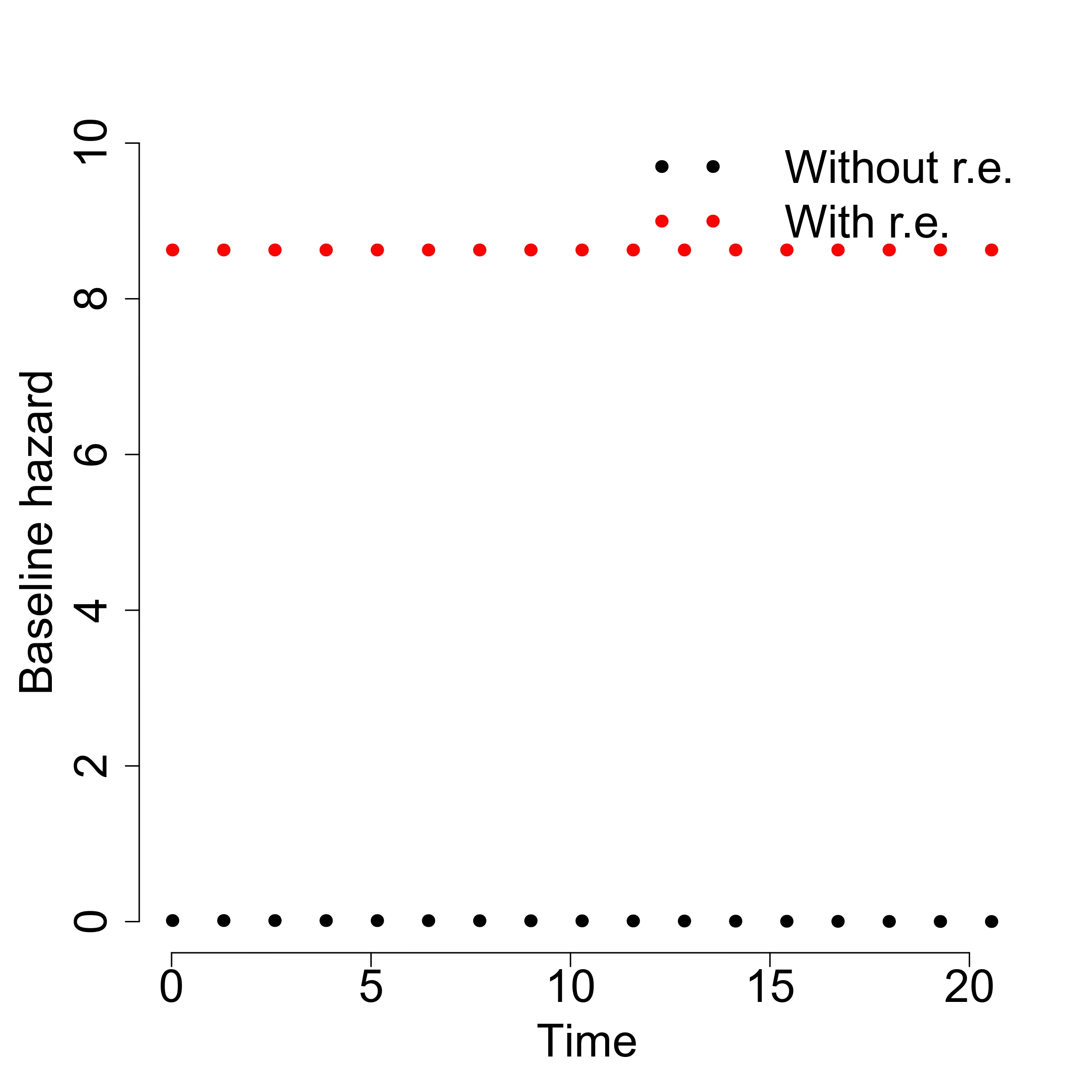}} & \adjustbox{valign=c}{\includegraphics[width=0.3\linewidth]{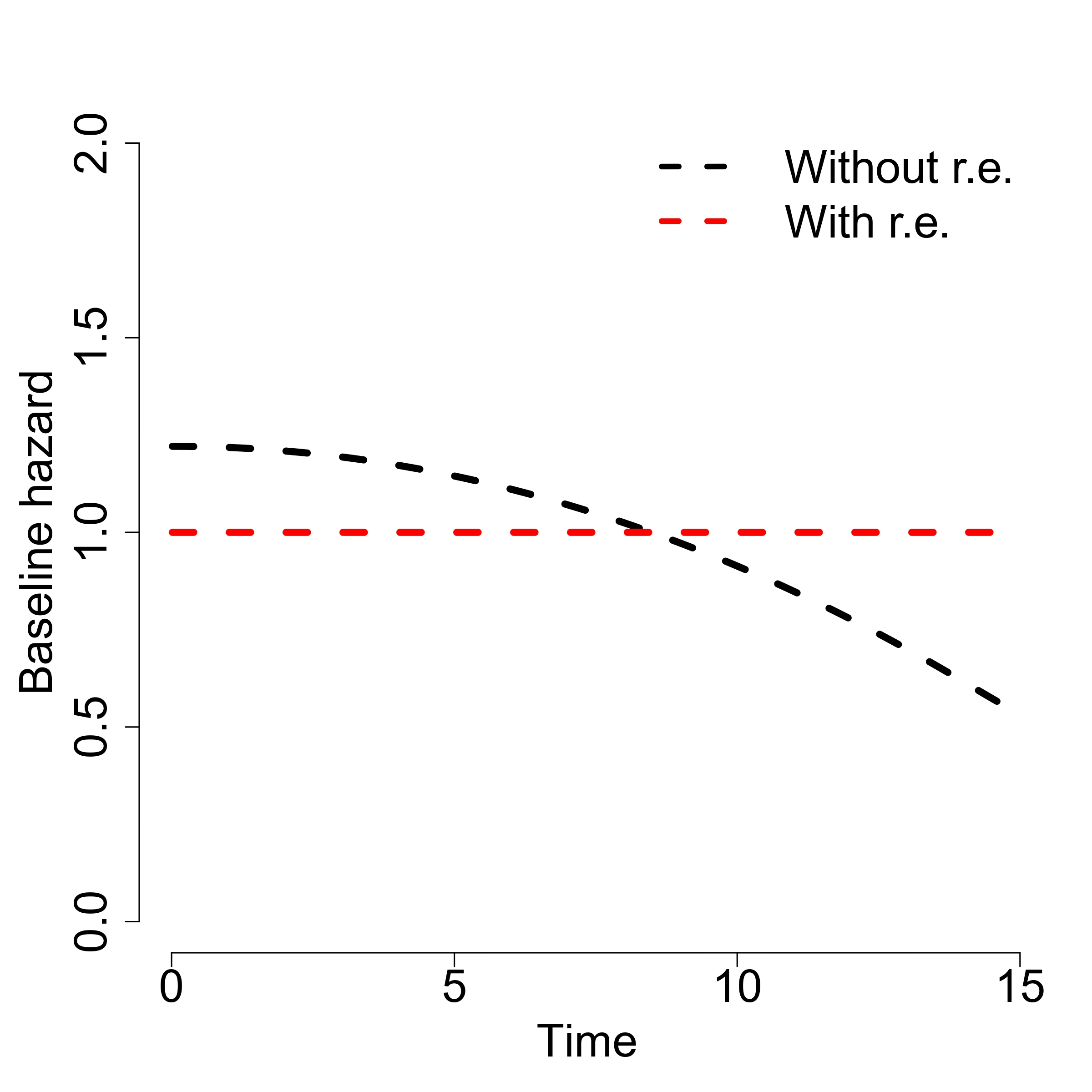}}\\
		\midrule
		VBS2 game & \adjustbox{valign=c}{\includegraphics[width=0.3\linewidth]{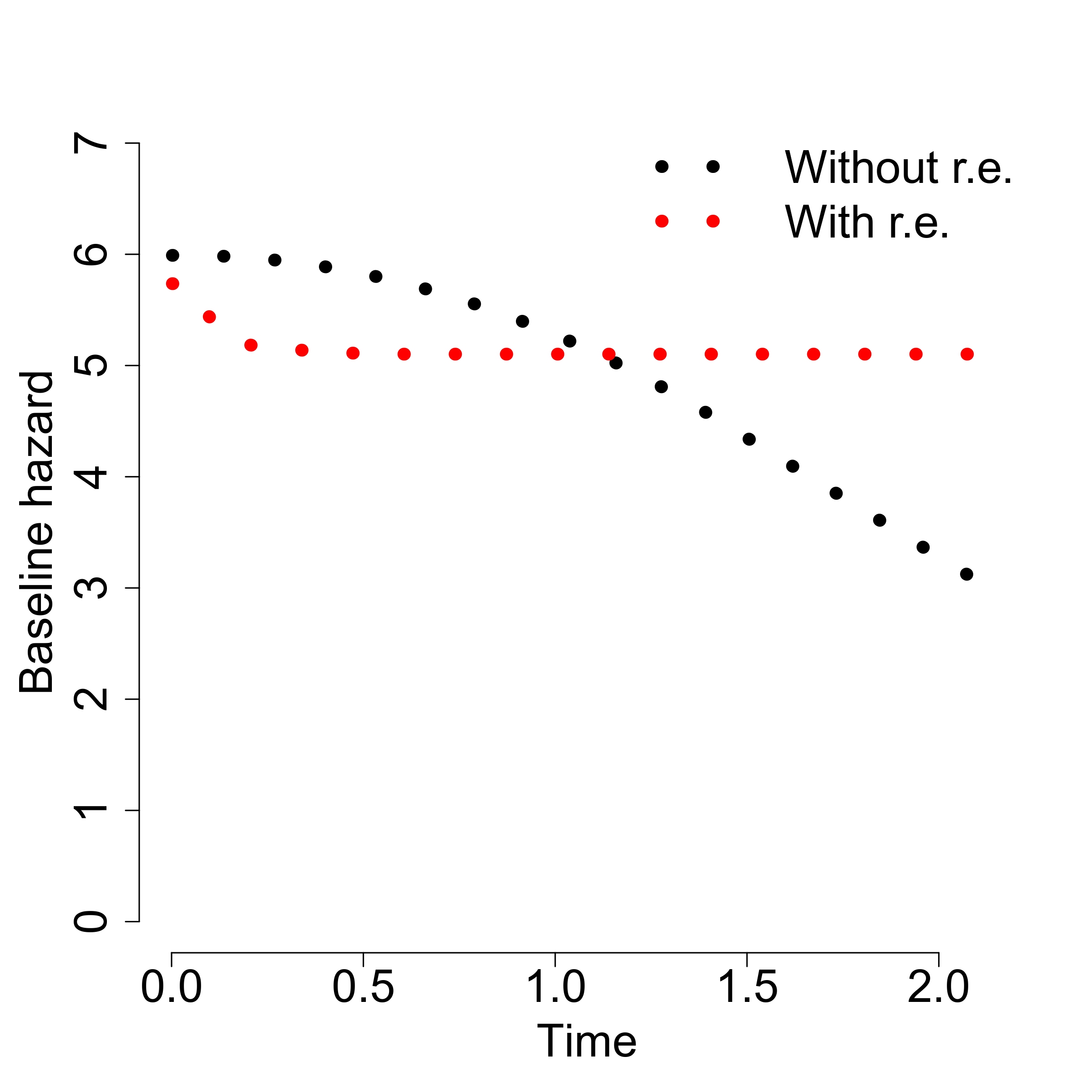}} & \\
	\end{longtable}

Table~\ref{tab:res} reports the model estimates for all datasets, including the result of the likelihood ratio test comparing the model with random effects to the model without random effect. Importantly, for 5 out of 6 examples analyzed we find that the random effects are not only statistically significant, but also with substantial standard deviations. This is on top of various nodal degree-based statistics that are included by default. Only in the VBS2 game data is there little evidence for nodal heterogeneity, possibly due to the fact that there are only 4 nodes present. This demonstrates that the proposed approach combined with a likelihood ratio test is capable of evaluating whether the inclusion of random effects enhances model performance. It does not advocate for the blind inclusion of the frailty terms in all cases.

\begin{table}[tb!]
	\fontsize{8pt}{9pt}\selectfont
	\centering
	\begin{tabular}{ccccccccccc}
		\toprule
		& $s_{\mbox{out}}$ & $s_{\mbox{in}}$ &$r_{\mbox{out}}$ &$r_{\mbox{in}}$  & rep & $t_{\mbox{turn}}$ & $c_{\mbox{turn}}$ & $\sigma_{pop}$ & $\sigma_{exp}$ & LRT \\
		\midrule
		\textbf{Manufacturing} & & & & & & & & & &\\
		without r.e.  & 0.08* & -0.04* & -0.05* & 0.13* & -0.002  & 0.005* & 0.002 & &  &  \\
		with r.e.  &  0.02* & 0.003 & 0.06* &0.001 & -0.006* & 0.004* & -0.002 & 1.6 & 1.44  & 0.00 \\
		\hline
		\textbf{Enron email} & & & & & & & & & &\\
		without r.e.  & 0.07* & -0.03 & 0.09* & -0.03 & 0.9*  & -1.2* & 14.43* & &  &  \\
		with r.e.  &  -0.01& -0.19*& 0.08*&-.78* & 1.55* & -1.62* & 6.75* & 0.9 & 1.59  & 0.00 \\
		\hline
		\textbf{Classroom} & & & & & & & & & &\\
		without r.e.  & -0.25*  &  -0.54*  &  -0.66* &-2.09* &15.06* &7.61* &$2.96^{\bigcdot}$ & & &\\
		with r.e.  & -0.29*  &  -1.43*  &  -0.54* &-1.87* &15.30* &11.28* &0.05 &0.41 &0.78 & 0.00\\
		\hline
		\textbf{Phone calls} & & & & & & & & & &\\
		without r.e.  & -0.08 &  0.46* & -0.31* & -0.39* & 2.99* & -2.25* & $-0.95^{\bigcdot}$ &  &  & \\
		with r.e. & -0.09 & -0.03 & -0.22* & -4.78* & 7.95* & -1.75* & 0.47 & 1.49 & 0.25 & 0.00\\
		\hline
		\textbf{Social evolution} & & & & & & & & & &\\
		without r.e.  &-$0.38^{\bigcdot}$ & 0.27 &-0.76* & -0.44 & 1.98* & -7.46* &  5.02* &  & &\\
		with r.e.  &  -1.49* &1.07* & -1.03* & -1.89* &3.77* & -10.19* & 1.28 &1.31 & 1.99&0.00\\
		\hline
		\textbf{VBS2 game} & & & & & & & & & &\\
		without r.e.  & 0.66 &-0.50 &-$5\cdot10^{-4}$ &-0.33 & 0.28* & 9.40 & -12.80 & & &\\
		with r.e.  & 0.66 &-0.50 &-$6\cdot10^{-4}$ &-0.33 & 0.28* & 9.40 & -12.80 & $2\cdot10^{-3}$ & $2\cdot10^{-3}$&1\\
		\bottomrule
	\end{tabular}
	\caption{Parameter estimates for the fixed and random effect standard
		deviations across all dataset (* p-value $<$ 0.05, $^{\bigcdot}$ p-value $<$ 0.1). The results of the partial likelihood-ratio test indicate a significant preference for the model with nodal random effects in the majority of cases. Only for the VBS2 game dataset, the model without random effects is found to be sufficient.}
	\label{tab:res}
\end{table}

We also provide significant test for all the degree based statistics. A number of them are significant, even in the presence of the random effects. This suggest that there is some level of \emph{emergence} or virality in these dynamic networks. For instance, in the Manufacturing company data coefficients associated with the sender out-degree and receiver out-degree statistics suggest that the mere fact of having communicated or having been contacted in the past makes a person more likely to initiate or receive emails in the future, respectively. Nevertheless, both effects are quite small compared to the standard deviation of the expansiveness and popularity frailty terms. This means that intrinsic nodal heterogeneity is more important, than the viral effects. We also observe a modest positive turn-taking effect, indicating that individuals tend to continue the discussion thread initiated by others. Moreover, this model suggest that repetition has a small negative effect in email communication, indicating that the past events have a reduced tendency to be repeated in the future. This phenomenon can be attributed to various factors such as evolving topics, shifting communication needs, or the dynamic nature of conversations. As individuals engage in email exchanges, they naturally adapt their communication patterns, explore new topics, and respond to evolving circumstances. This dynamic nature of email communication may lead to a decreased likelihood of direct repetition of the same events over time. However, the effect sizes of the latter two effects are relatively small, suggesting that repetition and turn-taking are less prominent compared to other network effects.

We end the empirical analysis section, by a more careful inspection of the various reciprocity and triadic closure effects for each of the datasets. Table \ref{tab:rec-tri} displays how the reciprocity and triadic effects curves can be biased, when one fails to account for nodal heterogeneity.  For instance, in the Manufacturing company example, failing to account for nodal heterogeneity seems to suggest a strong triadic closure tendency that rapidly decays over time. A similar situation can also be observed in the Classroom data and in the Phone call data. However, analysing a corresponding curve from the frailty model, which is supported by the likelihood ratio test, we can conclude that after accounting for nodal heterogeneity this triadic closure affect disappears. For the Enron email data and the Social evolution data, this effect is not so pronounced. In fact, we are quite confident that a triadic closure effect is really present in the Enron email data.

\section{Conclusions}

Given the complexity of real-world processes, nodal heterogeneity is likely to exist in most empirical networks. Common approaches to address heterogeneity include introducing various exogenous and endogenous network statistics. This is entirely sensible. However, the available exogenous and endogenous variables might not capture all the complexities of the generative process. For this reason, we stress the importance of being able to include frailty terms in the relational event model in order to account for this residual heterogeneity. Moreover, we argue that failing to account for heterogeneity can seriously affect inference about the strength of various endogenous network effects. Particularly, our work suggests that heterogeneity results in severely biased estimates of triadic closure effects. These results suggest that heterogeneity may also play a role in estimation of more complex network effects, including assortative matrix effects, four-cycles, etc. In addition, the case study analysis revealed that heterogeneity may also affect simpler network mechanisms, such as reciprocity. In this paper, we proposed using a relational even model frailty approach as a flexible tool to model heterogeneity in the network that is not otherwise captured in the available covariates. We also showed that likelihood ratio tests can be used to test for the need of including such frailty terms. 

The theoretical benefits of accounting for nodal heterogeneity have been illustrated through a simulation study. Numerical experiments confirmed that a failure to account for nodal heterogeneity can vastly overstate the prevalence of the triadic closure effect present in the network. Furthermore, the findings revealed that the transitive closure effect, due to its nature, is the most sensitive to the nodal heterogeneity. Additionally, the nature of the structural effect is closely related to the strength of influence of expansiveness and popularity. These nodal random effects depending on the type of triadic effect have a different impact on the emergence of ghost triadic closure effects. Simulation studies also confirmed that the frailty approach is capable of producing accurate estimates of the underlying parameters. As expected, the precision of the standard deviation estimates increase with the increasing number of levels of the random effect, i.e., higher replication of the random effects results in more precise estimates. Assessing the performance of the partial likelihood-ratio test, we noted that 
the test produces only slightly conservative estimates, demonstrating a suitable way to test for the inclusion of the frailty terms. 

This work revealed that nodal heterogeneity might disguise itself as a triadic closure effect when heterogeneity is not accounted for, i.e., if they are not well explained by the observed nodal characteristics. The suggested frailty approach is capable of recovering the temporal reciprocity and triadic closure curves, disentangling random nodal effects from triadic closure. The computational cost of the model is that of any traditional mixed effect model, making it possible to incorporate nodal random effects standardly in empirical studies.  

\section*{Acknowledgements}
This work was supported by an STSM Grant from COST Action COSTNET (CA15109). EW acknowledges funding by SNSF (grants 188534, 192549).

\vspace{\fill}\pagebreak

\appendix
\renewcommand{\theequation}{A\arabic{equation}}
\setcounter{equation}{0}
\renewcommand{\thesection}{\Alph{subsection}}
\setcounter{section}{0}

\bibliographystyle{chicago}
\bibliography{bibliography}

\begin{thebibliography}{}

\bibitem[\protect\citeauthoryear{Artico and Wit}{Artico and
  Wit}{2023}]{Igor2023}
Artico, I. and E.~C. Wit (2023, 04).
\newblock {Dynamic latent space relational event model}.
\newblock {\em Journal of the Royal Statistical Society Series A: Statistics in
  Society\/}~{\em 186\/}(3), 508--529.

\bibitem[\protect\citeauthoryear{Back}{Back}{2015}]{back2015opening}
Back, M.~D. (2015).
\newblock Opening the process black box: Mechanisms underlying the social
  consequences of personality.

\bibitem[\protect\citeauthoryear{Bianconi, Darst, Iacovacci, and
  Fortunato}{Bianconi et~al.}{2014}]{model_added_nodes}
Bianconi, G., R.-K. Darst, J.~Iacovacci, and S.~Fortunato (2014).
\newblock Triadic closure as a basic generating mechanism of communities in
  complex networks.
\newblock {\em Physical Review E\/}~{\em 90}.

\bibitem[\protect\citeauthoryear{Borgatti and Halgin}{Borgatti and
  Halgin}{2011}]{borgatti2011network}
Borgatti, S.~P. and D.~S. Halgin (2011).
\newblock On network theory.
\newblock {\em Organization science\/}~{\em 22\/}(5), 1168--1181.

\bibitem[\protect\citeauthoryear{Box-Steffensmeier, Campbell, Christenson, and
  Morgan}{Box-Steffensmeier et~al.}{2019}]{BoxSteffensmeier2019}
Box-Steffensmeier, J.~M., B.~W. Campbell, D.~Christenson, and J.~Morgan (2019).
\newblock Substantive implications of unobserved heterogeneity: Testing the
  frailty approach to exponential random graph models.
\newblock {\em Soc. Networks\/}~{\em 59}, 141--153.

\bibitem[\protect\citeauthoryear{Box-Steffensmeier, Christenson, and
  Morgan}{Box-Steffensmeier et~al.}{2018}]{box2018modeling}
Box-Steffensmeier, J.~M., D.~P. Christenson, and J.~W. Morgan (2018).
\newblock Modeling unobserved heterogeneity in social networks with the frailty
  exponential random graph model.
\newblock {\em Political Analysis\/}~{\em 26\/}(1), 3--19.

\bibitem[\protect\citeauthoryear{Butts}{Butts}{2008}]{butts}
Butts, C.-T. (2008).
\newblock A relational event framework for social action.
\newblock {\em Sociological Methodology\/}~{\em 38\/}(1), 155--200.

\bibitem[\protect\citeauthoryear{Butts, Lomi, Snijders, and Stadtfeld}{Butts
  et~al.}{2023}]{butts2023relational}
Butts, C.~T., A.~Lomi, T.~A. Snijders, and C.~Stadtfeld (2023).
\newblock Relational event models in network science.
\newblock {\em Network Science\/}~{\em 11\/}(2), 175--183.

\bibitem[\protect\citeauthoryear{Corbo, Corrado, and Ferriani}{Corbo
  et~al.}{2016}]{corbo2016new}
Corbo, L., R.~Corrado, and S.~Ferriani (2016).
\newblock A new order of things: Network mechanisms of field evolution in the
  aftermath of an exogenous shock.
\newblock {\em Organization Studies\/}~{\em 37\/}(3), 323--348.

\bibitem[\protect\citeauthoryear{DuBois, Butts, and Smyth}{DuBois
  et~al.}{2013}]{dubois2013stochastic}
DuBois, C., C.~Butts, and P.~Smyth (2013).
\newblock Stochastic blockmodeling of relational event dynamics.
\newblock In {\em Artificial intelligence and statistics}, pp.\  238--246.
  PMLR.

\bibitem[\protect\citeauthoryear{Fischbacher, G{\"a}chter, and
  Fehr}{Fischbacher et~al.}{2001}]{fischbacher2001people}
Fischbacher, U., S.~G{\"a}chter, and E.~Fehr (2001).
\newblock Are people conditionally cooperative? evidence from a public goods
  experiment.
\newblock {\em Economics letters\/}~{\em 71\/}(3), 397--404.

\bibitem[\protect\citeauthoryear{Foster, Foster, Grassberger, and
  Paczuski}{Foster et~al.}{2011}]{foster}
Foster, D.-V., J.-G. Foster, P.~Grassberger, and M.~Paczuski (2011).
\newblock Clustering drives assortativity and community structure in ensembles
  of networks.
\newblock {\em Physical Review E\/}~{\em 84}.

\bibitem[\protect\citeauthoryear{Gelman and Hill}{Gelman and
  Hill}{2006}]{gelman2006data}
Gelman, A. and J.~Hill (2006).
\newblock {\em Data analysis using regression and multilevel/hierarchical
  models}.
\newblock Cambridge university press.

\bibitem[\protect\citeauthoryear{Geukes, Breil, Hutteman, Nestler, K{\"u}fner,
  and Back}{Geukes et~al.}{2019}]{geukes2019explaining}
Geukes, K., S.~M. Breil, R.~Hutteman, S.~Nestler, A.~C. K{\"u}fner, and M.~D.
  Back (2019).
\newblock Explaining the longitudinal interplay of personality and social
  relationships in the laboratory and in the field: The pils and the connect
  study.
\newblock {\em PloS one\/}~{\em 14\/}(1), e0210424.

\bibitem[\protect\citeauthoryear{Hinde}{Hinde}{1979}]{hinde}
Hinde, R.~A. (1979).
\newblock {\em Towards understanding relationships}.
\newblock Published in cooperation with European Association of Experimental
  Social Psychology by Academic Press.

\bibitem[\protect\citeauthoryear{Isen}{Isen}{1987}]{isen1987positive}
Isen, A.~M. (1987).
\newblock Positive affect, cognitive processes, and social behavior.
\newblock {\em Advances in experimental social psychology\/}~{\em 20},
  203--253.

\bibitem[\protect\citeauthoryear{Juozaitienė and Wit}{Juozaitienė and
  Wit}{2022}]{JUOZAITIENE2022296}
Juozaitienė, R. and E.~C. Wit (2022).
\newblock Non-parametric estimation of reciprocity and triadic effects in
  relational event networks.
\newblock {\em Social Networks\/}~{\em 68}, 296--305.

\bibitem[\protect\citeauthoryear{Kevork and Kauermann}{Kevork and
  Kauermann}{2021}]{kevork2021iterative}
Kevork, S. and G.~Kauermann (2021).
\newblock Iterative estimation of mixed exponential random graph models with
  nodal random effects.
\newblock {\em Network Science\/}~{\em 9\/}(4), 478--498.

\bibitem[\protect\citeauthoryear{Klimek and Thurner}{Klimek and
  Thurner}{2013}]{Klimek}
Klimek, P. and S.~Thurner (2013, 01).
\newblock Triadic closure dynamics drives scaling-laws in social multiplex
  networks.
\newblock {\em New Journal of Physics\/}~{\em 15}.

\bibitem[\protect\citeauthoryear{Klimt and Yang}{Klimt and
  Yang}{2004}]{Klimt2004TheEC}
Klimt, B. and Y.~Yang (2004).
\newblock The enron corpus: A new dataset for email classification research.
\newblock In {\em European Conference on Machine Learning}, pp.\  217--226.

\bibitem[\protect\citeauthoryear{Kumpula, Onnela, Saram{\"a}ki, Kaski, and
  Kert{\'e}sz}{Kumpula et~al.}{2007}]{PhysRevLett}
Kumpula, J.~M., J.-P. Onnela, J.~Saram{\"a}ki, K.~Kaski, and J.~Kert{\'e}sz
  (2007).
\newblock Emergence of communities in weighted networks.
\newblock {\em Physical review letters\/}~{\em 99\/}(22), 228701.

\bibitem[\protect\citeauthoryear{Lerner, Bussmann, Snijders, and
  Brandes}{Lerner et~al.}{2013}]{lerner2013modeling}
Lerner, J., M.~Bussmann, T.~A. Snijders, and U.~Brandes (2013).
\newblock Modeling frequency and type of interaction in event networks.
\newblock {\em Corvinus journal of sociology and social policy\/}~{\em 4\/}(1),
  3--32.

\bibitem[\protect\citeauthoryear{Leskovec, Backstrom, Kumar, and
  Tomkins}{Leskovec et~al.}{2008}]{Leskovec:2008}
Leskovec, J., L.~Backstrom, R.~Kumar, and A.~Tomkins (2008).
\newblock Microscopic evolution of social networks.
\newblock In {\em Proceedings of the 14th ACM SIGKDD International Conference
  on Knowledge Discovery and Data Mining}, KDD '08, New York, NY, USA, pp.\
  462--470. ACM.

\bibitem[\protect\citeauthoryear{Li, Zou, Guan, Gong, Li, Di, and Lai}{Li
  et~al.}{2013}]{Li}
Li, M., H.~Zou, S.~Guan, X.~Gong, K.~Li, Z.~Di, and C.~Lai (2013, 08).
\newblock A coevolving model based on preferential triadic closure for social
  media networks.
\newblock {\em Scientific reports\/}~{\em 3}, 2512.

\bibitem[\protect\citeauthoryear{Lusher, Koskinen, and Robins}{Lusher
  et~al.}{2013}]{lusher2013exponential}
Lusher, D., J.~Koskinen, and G.~Robins (2013).
\newblock {\em Exponential random graph models for social networks: Theory,
  methods, and applications}.
\newblock Cambridge University Press.

\bibitem[\protect\citeauthoryear{Madan, Cebrian, Moturu, Farrahi, et~al.}{Madan
  et~al.}{2011}]{madan2011sensing}
Madan, A., M.~Cebrian, S.~Moturu, K.~Farrahi, et~al. (2011).
\newblock Sensing the" health state" of a community.
\newblock {\em IEEE Pervasive Computing\/}~{\em 11\/}(4), 36--45.

\bibitem[\protect\citeauthoryear{Mcfarland}{Mcfarland}{2001}]{Mcfarland}
Mcfarland, D. (2001, 11).
\newblock Student resistance: How the formal and informal organization of
  classrooms facilitate everyday forms of student defiance.
\newblock {\em American Journal of Sociology - AMER J SOCIOL\/}~{\em 107},
  612--678.

\bibitem[\protect\citeauthoryear{McPherson, Smith-Lovin, and Cook}{McPherson
  et~al.}{2001}]{mcpherson2001birds}
McPherson, M., L.~Smith-Lovin, and J.~M. Cook (2001).
\newblock Birds of a feather: Homophily in social networks.
\newblock {\em Annual review of sociology\/}~{\em 27\/}(1), 415--444.

\bibitem[\protect\citeauthoryear{Michalski, Kajdanowicz, Br{\'o}dka, and
  Kazienko}{Michalski et~al.}{2014}]{manufacturing_email}
Michalski, R., T.~Kajdanowicz, P.~Br{\'o}dka, and P.~Kazienko (2014).
\newblock Seed selection for spread of influence in social networks: Temporal
  vs. static approach.
\newblock {\em New Generation Computing\/}~{\em 32\/}(3-4), 213--235.

\bibitem[\protect\citeauthoryear{Newman and Park}{Newman and
  Park}{2003}]{PhysRevE}
Newman, M.-E.-J. and J.~Park (2003).
\newblock Why social networks are different from other types of networks.
\newblock {\em Physical Review E\/}~{\em 68}.

\bibitem[\protect\citeauthoryear{Olk and Gibbons}{Olk and
  Gibbons}{2010}]{olk2010dynamics}
Olk, P.~M. and D.~E. Gibbons (2010).
\newblock Dynamics of friendship reciprocity among professional adults.
\newblock {\em Journal of Applied Social Psychology\/}~{\em 40\/}(5),
  1146--1171.

\bibitem[\protect\citeauthoryear{Perry and Wolfe}{Perry and
  Wolfe}{2013}]{Wolfe}
Perry, P. and P.~Wolfe (2013, 11).
\newblock Point process modeling for directed interaction networks.
\newblock {\em Journal Of The Royal Statistical Society\/}~{\em 75\/}(5),
  821–849.

\bibitem[\protect\citeauthoryear{Pfeiffer, Rutte, Killingback, Taborsky, and
  Bonhoeffer}{Pfeiffer et~al.}{2005}]{pfeiffer2005evolution}
Pfeiffer, T., C.~Rutte, T.~Killingback, M.~Taborsky, and S.~Bonhoeffer (2005).
\newblock Evolution of cooperation by generalized reciprocity.
\newblock {\em Proceedings of the Royal Society B: Biological Sciences\/}~{\em
  272\/}(1568), 1115--1120.

\bibitem[\protect\citeauthoryear{Pilny, Schecter, Poole, and Contractor}{Pilny
  et~al.}{2016}]{pilny2016illustration}
Pilny, A., A.~Schecter, M.~S. Poole, and N.~Contractor (2016).
\newblock An illustration of the relational event model to analyze group
  interaction processes.
\newblock {\em Group Dynamics: Theory, Research, and Practice\/}~{\em 20\/}(3),
  181--195.

\bibitem[\protect\citeauthoryear{Raush}{Raush}{1965}]{raush1965interaction}
Raush, H.~L. (1965).
\newblock Interaction sequences.
\newblock {\em Journal of personality and social psychology\/}~{\em 2\/}(4),
  487.

\bibitem[\protect\citeauthoryear{Robins, Pattison, and Wang}{Robins
  et~al.}{2009}]{robins2009closure}
Robins, G., P.~Pattison, and P.~Wang (2009).
\newblock Closure, connectivity and degree distributions: Exponential random
  graph (p*) models for directed social networks.
\newblock {\em Social Networks\/}~{\em 31\/}(2), 105--117.

\bibitem[\protect\citeauthoryear{Rutte and Taborsky}{Rutte and
  Taborsky}{2007}]{rutte2007generalized}
Rutte, C. and M.~Taborsky (2007).
\newblock Generalized reciprocity in rats.
\newblock {\em PLoS Biology\/}~{\em 5\/}(7), e196.

\bibitem[\protect\citeauthoryear{Sapiezynski, Stopczynski, Lassen, and
  Lehmann}{Sapiezynski et~al.}{2019}]{sapiezynski2019interaction}
Sapiezynski, P., A.~Stopczynski, D.~D. Lassen, and S.~Lehmann (2019).
\newblock Interaction data from the copenhagen networks study.
\newblock {\em Scientific Data\/}~{\em 6\/}(1), 315.

\bibitem[\protect\citeauthoryear{Snijders, {van de Bunt}, and
  Steglich}{Snijders et~al.}{2010}]{Snijders2010}
Snijders, T., G.~{van de Bunt}, and C.~Steglich (2010).
\newblock Introduction to stochastic actor-based models for network dynamics.
\newblock {\em Social Networks\/}~{\em 32}, 44--60.

\bibitem[\protect\citeauthoryear{Snijders}{Snijders}{2017}]{snijders2017stochastic}
Snijders, T.~A. (2017).
\newblock Stochastic actor-oriented models for network dynamics.
\newblock {\em Annual review of statistics and its application\/}~{\em 4},
  343--363.

\bibitem[\protect\citeauthoryear{Stadtfeld and Block}{Stadtfeld and
  Block}{2017}]{Stadtfeld_2017}
Stadtfeld, C. and P.~Block (2017).
\newblock Interactions, actors, and time: Dynamic network actor models for
  relational events.
\newblock {\em Sociological Science\/}~{\em 4\/}(14), 318--352.

\bibitem[\protect\citeauthoryear{Steele}{Steele}{2003}]{steele2003discrete}
Steele, F. (2003).
\newblock A discrete-time multilevel mixture model for event history data with
  long-term survivors, with an application to an analysis of contraceptive
  sterilization in bangladesh.
\newblock {\em Lifetime Data Analysis\/}~{\em 9\/}(2), 155--174.

\bibitem[\protect\citeauthoryear{Therneau and Grambsch}{Therneau and
  Grambsch}{2000}]{survival-book}
Therneau, T. and P.~Grambsch (2000).
\newblock {\em Modeling Survival Data: Extending the {C}ox Model}.
\newblock New York: Springer.

\bibitem[\protect\citeauthoryear{Thiemichen, Friel, Caimo, and
  Kauermann}{Thiemichen et~al.}{2016}]{thiemichen2016}
Thiemichen, S., N.~Friel, A.~Caimo, and G.~Kauermann (2016).
\newblock Bayesian exponential random graph models with nodal random effects.
\newblock {\em Social Networks\/}~{\em 46}, 11--28.

\bibitem[\protect\citeauthoryear{Uzaheta, Amati, and Stadtfeld}{Uzaheta
  et~al.}{2023}]{uzaheta2023random}
Uzaheta, A., V.~Amati, and C.~Stadtfeld (2023).
\newblock Random effects in dynamic network actor models.
\newblock {\em Network Science\/}~{\em 11\/}(2), 249--266.

\bibitem[\protect\citeauthoryear{Vu, Lomi, Mascia, and Pallotti}{Vu
  et~al.}{2017}]{vu2017relational}
Vu, D., A.~Lomi, D.~Mascia, and F.~Pallotti (2017).
\newblock Relational event models for longitudinal network data with an
  application to interhospital patient transfers.
\newblock {\em Statistics in medicine\/}~{\em 36\/}(14), 2265--2287.

\bibitem[\protect\citeauthoryear{Yarmoshuk, Cole, Mwangu, Guantai, and
  Zarowsky}{Yarmoshuk et~al.}{2020}]{yarmoshuk2020reciprocity}
Yarmoshuk, A.~N., D.~C. Cole, M.~Mwangu, A.~N. Guantai, and C.~Zarowsky (2020).
\newblock Reciprocity in international interuniversity global health
  partnerships.
\newblock {\em Higher Education\/}~{\em 79\/}(3), 395--414.

\end{thebibliography}

\newpage

\section*{Social evolution study: model output}
The final random effects model selected also included two fixed effects, namely whether they lived on the same floor and whether they were in the same year. Both effects are positive, suggesting that sharing floor and year increases the rate of interaction. 

\begin{Verbatim}[baselinestretch=1]
		coef     exp(coef)   se(coef)    z     p
s_in         1.0730086 2.924164e+00 0.3073203  3.49 4.8e-04
r_out       -1.0296574 3.571293e-01 0.2860164 -3.60 3.2e-04
s_out       -1.4869406 2.260632e-01 0.2970077 -5.01 5.5e-07
r_in        -1.8933611 1.505649e-01 0.4314381 -4.39 1.1e-05
rep          3.7738430 4.354710e+01 0.3963549  9.52 0.0e+00
turntaking -10.1903718 3.752993e-05 1.9613216 -5.20 2.0e-07
turnconti    1.2800694 3.596889e+00 1.7870746  0.72 4.7e-01
floor        0.7264987 2.067828e+00 0.1997797  3.64 2.8e-04
grade        0.7492175 2.115344e+00 0.1987526  3.77 1.6e-04
	
Random effects
Group    Variable  Std Dev  Variance
sender   Intercept 1.989077 3.956427
receiver Intercept 1.313375 1.724953
\end{Verbatim}

\section*{Classroom study: model output}
The final random effects model selected also included three fixed effects, namely whether the receiver is female, whether the sender is a teacher and whether the receiver is a teacher. The first effect is not significant, whereas teachers have higher sending propensity and a lower receiving propensity, compared to students. 

\begin{Verbatim}[baselinestretch=1]
	                coef       exp(coef)   se(coef)    z       p
s_out               -0.29373674  7.454727e-01 0.04857225 -6.05  1.5e-09
s_in                -1.43426092  2.382914e-01 0.23616393 -6.07  1.3e-09
r_out               -0.53721881  5.843712e-01 0.10828762 -4.96  7.0e-07
r_in                -1.87499243  1.533561e-01 0.24265821 -7.73  1.1e-14
rep                  15.30071436 4.415865e+06 0.72145952  21.21 0.0e+00
turntaking           11.28109797 7.930829e+04 2.02154071  5.58  2.4e-08
turnconti            0.04634697  1.047438e+00 1.85815115  0.02  9.8e-01
Receiver is Female   0.26393685  1.302046e+00 0.21889317  1.21  2.3e-01
Sender is Teacher    2.26734670  9.653752e+00 0.59845563  3.79  1.5e-04
Receiver is Teacher -1.15939004  3.136775e-01 0.40643327 -2.85  4.3e-03
	
Random effects
Group    Variable  Std Dev   Variance 
sender   Intercept 0.7757925 0.6018541
receiver Intercept 0.4091359 0.1673922
\end{Verbatim}



%




\end{document}